\documentclass{aa}  

\usepackage{graphicx}
\usepackage{txfonts}
\usepackage{color}
\usepackage{soul}

\begin{document}

\title{Exploring the physics behind the non-thermal emission from star-forming galaxies detected in gamma rays}
\titlerunning{Non-thermal emission from gamma-ray star-forming galaxies}

\author{P. Kornecki\inst{1}, E. Peretti\inst{2}, S. del Palacio\inst{1}, P. Benaglia\inst{1}, L.~J. Pellizza\inst{3}}

\institute{Instituto Argentino de Radioastronom\'{\i}a, CONICET-CICPBA-UNLP, CC5 (1897) Villa Elisa, Prov. de Buenos Aires, Argentina \and Niels Bohr International Academy, Niels Bohr Institute, University of Copenhagen, Blegdamsvej 17, DK-2100 Copenhagen, Denmark \and
Instituto de Astronomía y Física del Espacio, CONICET-UBA, C.C. 67, Suc. 28, 1428 Buenos Aires, Argentina }

   \date{}
 
  \abstract
   {Star-forming galaxies emit non-thermal radiation from radio to $\gamma$-rays. Observations show that their radio and $\gamma$-ray luminosities scale with their star formation rates, supporting the hypothesis that non-thermal radiation is emitted by cosmic rays produced by their stellar populations. However, the nature of the main cosmic-ray transport processes that shape the emission in these galaxies is still poorly understood, especially at low star formation rates.}
   {We aim to investigate the main mechanisms of global cosmic-ray transport and cooling in star-forming galaxies. The way they contribute in shaping the relations between non-thermal luminosities and star formation rates could shed light onto their nature, and allow to quantify their relative importance at different star formation rates.}
   {We develop a model to compute the cosmic-ray populations of star-forming galaxies, taking into account their production, transport and cooling. The model is parametrised only through global galaxy properties, and describes the non-thermal emission in both radio (at 1.4~GHz and 150~MHz) and $\gamma$-rays (in the 0.1--100~GeV band). We focus on the role of diffusive and advective transport by galactic winds, either driven by turbulent or thermal instabilities. We compare model predictions to observations, for which we compile a homogeneous set of luminosities in these radio bands, and update those available in $\gamma$ rays.}
   {Our model reproduces reasonably well the observed relations between the $\gamma$-ray or 1.4~GHz radio luminosities and the star formation rate, assuming a single power-law scaling of the magnetic field with the latter with index $\beta=0.3$, and winds blowing either at Alfvenic speeds ($\sim$ tens of km~s$^{-1}$, for $\lesssim 5\,\mathrm{M_\odot\,yr^{-1}}$) or typical starburst wind velocities ($\sim$ hundreds of km~s$^{-1}$, for $\gtrsim 5\,\mathrm{M_\odot\,yr^{-1}}$). Escape of cosmic rays is negligible for $\gtrsim 30\,\mathrm{M_\odot\,yr^{-1}}$. A constant ionisation fraction of the interstellar medium fails to reproduce the 150~MHz radio luminosity throughout the whole star formation rate range.}
   {Our results reinforce the idea that galaxies with high star formation rate are cosmic-ray calorimeters, and that the main mechanism driving proton escape is diffusion, whereas  electron escape also proceeds via wind advection. They also suggest that these winds should be cosmic-ray or thermally-driven at low and intermediate star formation rates, respectively. Our results globally support that magneto-hydrodynamic turbulence is responsible for the dependence of the magnetic field strength on the star formation rate and that the ionisation fraction is strongly disfavoured to be constant  throughout the whole range of star formation rate.}


\maketitle
%

\section{Introduction}

Star forming galaxies (SFGs) comprise one of the extragalactic $\gamma$-ray sources classes detected at GeV energies by the \textit{Fermi-LAT} telescope \citep{Abdo2010,Lenain2010,Ackermann2012, Abdo2010a}.
Their emission is believed to be associated with young stellar populations, which explains its correlation with the star formation rate (SFR).
In particular, $\gamma$-rays are thought to be mostly produced in interactions between hadronic cosmic-rays (CRs) with particles of the interstellar medium (ISM), while CRs are accelerated at supernova remnant (SNR) shocks \citep{Jokipii1985,Blasi-review,2017ApJ...835...72B}.
Beside SNRs, other sources related to star forming regions such as young massive star clusters or microquasars could also play an important role as CR acceleration sites, possibly up to PeV energies \citep[see e.g.][]{Bykov2001,Aharonian2019,Bykov2020Rev,Escobar2021,Morlino2021}. 
These acceleration sites are abundant in SFGs and are expected to be more numerous in sources with higher SFR. 
Similarly, higher gas densities are found in higher SFR environments since the gas is the seed of star formation.
As a result, an increase in the $\gamma$-ray emission is expected for higher star formation rates.

The main evidence supporting the aforementioned interpretation is the observed correlation between the integrated $\gamma$-ray luminosity ($L_\gamma$) of SFGs and their SFR \citep{Ackermann2012, ajello2020}.
The latter, in particular, is typically estimated from tracers such as the total infrared (IR) luminosity ($L_\mathrm{IR}$, $8-1000\,\mu\mathrm{m}$).
This relation was predicted by \citet{Thompson2007} and \citet{Lacki2010} in terms of the $\gamma$-ray flux and the gas surface density of SFGs.
Several works have endeavoured to explain the high-energy emission of these objects individually \citep[e.g.,][]{Volk1996, Blom1999, Romero2003, Domingo2005, deCae2009, persic2010, Lacki2011M82NGC253, Rephaeli2014, Krumholz2020} and others as a result of the common properties of the population of SFGs \citep[e.g.,][]{Lacki2010, Pfrommer2017, Zhang2019, peretti2020, Kornecki2020}. 
There is a general consensus on the increasing dynamical relevance of radiative energy losses for higher SFR, so that the CR transport in these environments is inferred to approach the calorimetric limit, namely all injected particles cool before being able to escape. 
In particular, while electron calorimetry is believed to be reached already in mild starbursts, SFR $\gtrsim 5 \, \rm M_{\odot} \, yr^{-1}$, the SFR range where proton calorimetry might be achieved is still under debate due to our current poor knowledge of the escape mechanisms.

Aside $\gamma$-rays, SFGs are also observed and modelled throughout the whole electromagnetic spectrum \citep[e.g.,][]{Volk1996, Persic2002, 2008A&A...486..143P, deCae2009, Rephaeli2010, Yoast-Hull_M82, Persic2014, Peretti2019}. 
Diffuse thermal and non-thermal emission from these objects has been observed in radio frequencies \citep[e.g.,][]{Condon1992, Cox1988, Loiseau1987, Strickland2002, Kapinska2017}, X-rays \citep{Bauer2008, Sarkar2016} and even very-high-energy $\gamma$-rays \citep{2009Natur.462..770V, 2009Sci...326.1080A}.

Non-thermal radio and X-rays are produced by relativistic primary and secondary CR electrons, where primaries are  injected together with protons at acceleration sites \citep[see e.g.][in the context of SNRs and star clusters respectively]{Cristofari2021,Aharonian2019} and secondaries are a by-product of the CR interaction with the ISM.
CR electrons in galactic magnetic fields ($B$) emit synchrotron radiation detectable mostly at radio frequencies. 
A connection between the synchrotron radiation and supernova rate was one of the first attempts to explain the linear correlation between the observed luminosity at 1.4~GHz ($L_{\rm 1.4 GHz}$) and far IR (FIR) luminosity 
\citep{van_der_ruit1971, Condon1992, Yun2001}.
However, the nature of this link is not trivial in low-SFR galaxies, due to their low efficiency in converting starlight into IR radiation, and the predominance of escape as the main driver of electron transport \citep{Condon1992, Yun2001, Bell2003}. This missing energy causes both radio and IR luminosities to underestimate the SFR for galaxies with low IR emission. \citet{Bell2003} argued that these two effects conspire to achieve a linear relation between IR luminosity and $L_{\rm 1.4 GHz}$. 
At high SFR, the efficient cooling of protons typical of starburst galaxies (SBGs) leads to a copious production of secondary leptons whose synchrotron emission is inferred to overcome that of primaries \citep[e.g.][]{Torres2004, Domingo2005, Lacki2010}. This could possibly impact  the global shape of the correlation.

In addition to synchrotron, also thermal processes such as free--free emission and absorption characterise the radio emission of SFGs.
In fact, numerous young stars create several H\textsc{ii} regions, whose hot ionised gas radiates thermally, contributing to the total radio emission. A spectral identification of such free--free emission is in general very complex due to the presence of other components such as synchrotron, dominating at GHz, and the FIR thermal peak produced by the dust, which arises above a few tens of GHz. 
Many works have discussed the fraction of the thermal contribution in such band with respect to the non-thermal emission \citep[e.g.][]{Bell2003, Klein2018}.
However, no definitive conclusion has been reached yet due to the complexity of this task.

Focusing on the synchrotron emission, many studies
explored the SFR--luminosity correlation at lower radio frequencies, in particular at 150~MHz \citep[e.g.][]{Gurkan2018,Wang2019,Smith2020}, 
where the free--free absorption might become severe.
Typically, the free-free absorption significantly reduces the emission up to $\sim 10-10^2$ MHz in galaxies with moderate SFR \citep{Schober2017,chyzy2018}, 
and up to a few GHz in ultraluminous IR galaxies \citep[ULIRG;][]{Condon1991, Clemens2010}. 

Despite most of the radio emission is expected from CRs populating galactic disks, some additional radio emission produced in SFGs might have different origin.
In particular, a non--negligible contribution might come from individual sources such as SNRs \citep[][]{Lisenfeld2000}, activity of a coexisting active galactic nucleus  \citep[AGN;][]{Mancuso2017}, or particles populating the galactic halo as a consequence of the escape from the galactic disk via diffusion or advection in winds \citep{Heesen2009, Romero2018}.
It is also possible a smaller contribution of emission coming from pulsar wind nebulae \citep{Ohm2013} and millisecond pulsars \citep{Sudoh2021}; in particular, the latter can become important at low frequencies and low SFR. Nevertheless, the contribution of these objects to the diffuse emission is still unclear.

\citet{Lacki2010} and \citet{Martin2014} have explored the transport of CRs throughout the SFR range, studying at the same time both the radio and $\gamma$-ray correlations with the SFR. 
This multiwavelength approach allowed to better constrain the role of CRs in shaping these correlations highlighting, in particular, the key role played by protons in the production of both $\gamma$-rays and secondary electrons. Indeed, the latter are inferred to be responsible for most of the radio emission at high SFR, where protons are expected to be efficiently confined.
Despite these recent improvements, our knowledge of the CR transport properties in SFGs remains poor and several questions are still open. For instance: 
How does the CR escape from galaxies evolve with the SFR and which is its impact on calorimetry?
What is the interplay between primary and secondary electrons in shaping the radio luminosity of SFGs?
Can the luminosity--SFR correlation help us in exploring possible dependencies of magnetic and radiation fields in SFGs? What is the actual role/contamination of free--free emission and absorption in SFGs? Can it help us in exploring alternative cosmic acceleration sites? 

In this work we address these issues by exploring the non-thermal and thermal emissions in galaxies detected in $\gamma$-rays, and their relation with the SFR at different bands (GeV $\gamma$ rays and 1.4~GHz and 150~MHz in radio).
We provide for the first time a nearly homogeneous 
set of observations, and compute the general trend for the $L_{\gamma}$, $L_{1.4 \rm HGz}$ and $L_{150 \rm MHz}$--SFR correlations for all SFGs detected in $\gamma$-rays so far. In particular, we update the $L_{\gamma}$--$L_{1.4 \rm HGz}$ correlation presented by \citet{Ackermann2012}.
In order to understand the nature of these correlations, and improving on the work of \citet[][hereafter K20]{Kornecki2020}, we develop a detailed model of the non-thermal emission taking into account also the thermal component produced by the ionised gas and the associated absorption.
Different from \citet{Lacki2010}, we allow for advection escape in the entire range of SFR. 
In addition, we attempt to capture the main properties of CR diffusion proposing a model for CR escape adequate to the source activity level, comprising SFR--dependent diffusion coefficient. 
Besides, we explore different scaling laws of $B$ with the SFR. 
Unlike \citet{Martin2014}, we focus on the singular contribution of primary and secondary electrons to the radio correlations. This allows us to explore the impact of calorimetric transport on such correlations. Additionally we include and discuss the role of free-free emission and absorption. 
We analyse carefully the effects of model parameters (as the total ionisation fraction of the ISM) and scaling relations needed to describe the emission at the three mentioned different wavelength.

The paper is organised as follows: In Section~\ref{Sec: obs_correl} we present the data set used and the luminosity--SFR correlations.
In Sections~\ref{Sec: model} and \ref{Sec: Model and Scale relations} we present the model and the scale relationships used, taking as a starting point the ones proposed in K20. 
In Section~\ref{Sec: Results} we present our main results, test the robustness of the model, and compare its predictions with observations to assess its reliability and delimit the most probable region of the parameter space. 
We also discuss our results and compare them with the literature in this section. Finally, in Section~\ref{Sec: conclusions} we present our conclusions.

%
\section{The Luminosity--SFR relation}
 \label{Sec: obs_correl}
%

We aim to address the physics behind the observed luminosity--SFR relations in a panchromatic fashion, specifically in the radio and $\gamma$-ray bands that trace non-thermal emission from relativistic particles. The existence of a tight correlation between the SFR of a galaxy and its luminosities in these bands ($L_{\gamma}$ and $L_\mathrm{radio}$, respectively) spanning more than four decades in SFR, indicates that non-thermal emission processes are common to all SFGs. The most compelling interpretation for these relations is that they arise from the emission of the population of cosmic rays within SFGs. Other scenarios such as the emission in jets or super-winds have been associated with non-thermal emission, but there is no evidence supporting that they are ubiquitous in SFGs.

Among these two bands, the sample of galaxies detected in $\gamma$-rays is the most restrictive. We therefore focus on this sample, that comprises 14 objects. These SFGs have a steady $\gamma$-ray emission \citep{Ackermann2012, ajello2020}, with the exception of the \object{Circinus Galaxy}, that shows marginal evidence of variability \citep{Guo2019}. This favours the interpretation that their high-energy emission is associated with their stellar populations instead of jets. We note that, due to their different sizes and distances, and also because of the different resolving power of $\gamma$-ray and radio telescopes, some of these galaxies are detected as point sources, whereas others are resolved. We do not take into account the structure of the resolved sources, because our main goal is to study the link between the diffuse emission of the galaxies and their global SFR.

In the following sections, we construct our sample by taking from the literature the observed fluxes of each of the 14 galaxies currently detected in $\gamma$ rays. We use the homogeneous set of distances and SFRs determined by K20 to calculate their radio and $\gamma$-ray luminosities in a self-consistent way. We also adopt from K20 the SFR values for the galaxies in our sample. These SFRs were computed with  $L_\mathrm{IR}$-independent methods when possible, in order to obtain a reliable SFR estimate for unobscured systems. In Table~\ref{tabledata} we present the data on distances, and fluxes and luminosities in the relevant bands, together with their respective errors and references.

\begin{table*}
    \caption{Distances, 1.4 GHz fluxes, 150 MHz fluxes,  and luminosities for all $\gamma$-ray emitting SFGs known.}
    \centering
    \begin{tabular}{lcccccc}
    \hline\hline
    Galaxy & $D_\mathrm{L}$ & $F_\mathrm{1.4\, GHz}$ & $F_\mathrm{150\, MHz}$ & $\log{(L_\gamma)}$ & $\log{(L_\mathrm{1.4\, GHz})}$ & $\log{(L_\mathrm{150\, MHz})}$  \\
    & [Mpc] & [mJy] & [mJy] &$\mathrm{erg\, s^{-1}}$ &  [$\mathrm{W/ Hz}$] &  [$\mathrm{W/ Hz}$] \\
    \hline 
    M31 & $0.77 \pm 0.04$\tablefootmark{a}  & $ 8600 \pm 1100$ \tablefootmark{b}  & $219000\pm 21000 $\tablefootmark{o} & $38.10 \pm 0.12$ & $ 21.28 \pm 0.14 $ & 22.19 $\pm$ 0.06\\
    NGC 253  & $3.56 \pm 0.26$\tablefootmark{a} & $5677 \pm 81 $ & $16600 \pm 1400$\tablefootmark{p} & $40.01 \pm 0.07$ & $21.93 \pm 0.06$ & $22.40 \pm 0.07$\\
    SMC      & $0.060 \pm 0.003$\tablefootmark{a} & $ 57000 \pm 5000 $ \tablefootmark{c} & $258300\pm	67300$\tablefootmark{q} & $37.04 \pm 0.05$ & $ 19.39 \pm 0.06$ & $20.05 \pm 0.12$\\
    M33      & $0.91 \pm 0.04$\tablefootmark{a} & $ 3300 \pm 500$ \tablefootmark{b} & $5600 \pm 	1800$\tablefootmark{u} & $38.30 \pm 0.09$ & $20.51 \pm 0.08$ & $20.74 \pm 0.14$\\
    NGC 1068          &$10.1 \pm 1.8$\tablefootmark{a} & $5027 \pm 7$ & $21230 \pm 3400$\tablefootmark{v} & $40.90 \pm 0.16$ & $21230 \pm 3400$ & $23.41 \pm 0.17$\\
    LMC      &$0.050 \pm 0.003$\tablefootmark{a}  & $ 426000\pm 29000$\tablefootmark{d} & $1450100\pm	246600$\tablefootmark{q} & $37.77 \pm 0.05$ & $20.10 \pm 0.06$ & $20.64 \pm 0.09$\\
    NGC 2146 & $17.2 \pm 3.2$\tablefootmark{a}& $1082 \pm 3 $ & $4040\pm 90$\tablefootmark{w} & $40.76 \pm 0.18$ & $22.58 \pm 0.16 $ & $23.15 \pm 0.16$\\
    NGC 2403 & $3.18 \pm 0.18$\tablefootmark{a} & $148 \pm 0.5 $ & $970 \pm 50$\tablefootmark{r} & $39.17 \pm 0.11$ & $20.25 \pm 0.05$ & $21.07 \pm 0.05 $\\
    M82   & $3.53 \pm 0.26$\tablefootmark{a}& $ 7389 \pm 13 $ & $15840 \pm 790$\tablefootmark{r} & $40.20 \pm 0.07$ & $22.04 \pm 0.06$ & $22.37 \pm 0.07$\\
    NGC 3424 & $25.6 \pm 1.8$\tablefootmark{a} & $ 71 \pm 1.8 $ & $230 \pm 30$\tablefootmark{r} & $40.92 \pm 0.13$ & $21.75 \pm 0.06$ & $22.26 \pm 0.08$\\
    Arp 299 & $46.8 \pm 3.3$\tablefootmark{a} & $688 \pm 60 $ & $4470 \pm 400$\tablefootmark{s} & $41.52 \pm 0.12$ & $23.25 \pm 0.07$ & $24.07 \pm 0.07 $\\
    NGC 4945 & $3.72 \pm 0.27$\tablefootmark{a} & $7953.52 \pm 133$ & $*$ & $40.27 \pm 0.07$ & $22.12 \pm 0.06$ & *\\
    Circinus & $4.21 \pm 0.70$\tablefootmark{a} & $1732.79 \pm 218$ & $*$ & $40.15 \pm 0.16$ & $21.56 \pm 0.15 $& *\\
    Arp 220 & $80.9 \pm 5.7$\tablefootmark{a} & $ 332 \pm 3$ & $315 \pm 59$\tablefootmark{t} & $42.41 \pm 0.08$ & $23.41 \pm 0.06 $ & $23.39 \pm 0.10$\\
    Milky Way  & *& *& $*$ & $38.91 \pm 0.13$\tablefootmark{n} & $21.28 \pm 0.14$\tablefootmark{n}& *\\
    \hline\hline
    \end{tabular}
    \tablefoot{
    \tablefoottext{a}{\citet{Kornecki2020} and references therein.}
    \tablefoottext{b}{\citet{Dennison1975}}
    \tablefoottext{c}{\citet{Loiseau1987}}
    \tablefoottext{d}{\citet{Hughes2007}}
    \tablefoottext{g}{\citet{4FGL2020}.}
    \tablefoottext{l}{Computed from $F_\mathrm{1.4\, GHz}$}
    \tablefoottext{n}{\citet{strong2010}}
    \tablefoottext{o}{\citet{Durdin1972}}
    \tablefoottext{p}{\citet{Kapinska2017}}
    \tablefoottext{q}{\citet{For2018}}
    \tablefoottext{r}{\citet{chyzy2018}}
    \tablefoottext{s}{\citet{Cox1988}}
    \tablefoottext{t}{\citet{Varenius2016}}
    \tablefoottext{u}{\citet{Israel1992}}
    \tablefoottext{v}{\citet{Kuehr1981}}
    \tablefoottext{w}{\citet{Hale1991}}
    }
    \label{tabledata}
\end{table*}

\subsection{The $L_\gamma$--SFR relation} \label{sec:Lgamma_vs_SFR}
We revisit the $L_\gamma$--SFR correlation studied by K20 and reconstruct it with the most recent $\gamma$-ray fluxes, taken from the Fermi Large Area Telescope Fourth Source Catalogue Data Release 2 \citep[4FGLR2,][]{4FGL2020R2}. This catalogue presents a new detection (\object{Arp 299}) compared to the previous release used by K20 \citep[4FGL,][]{4FGL2020}. 

We compute the $\gamma$-ray luminosities using the observed fluxes in the $0.1$--$100\,\mathrm{GeV}$ band and the luminosity distances $D_L$ from the self-consistent set reported in K20. For two objects, \object{M33} and \object{NGC 2403}, data in the $0.1$--$100\,\mathrm{GeV}$ band are not reported in 4FGLR2; we thus compute their fluxes from the $0.1$--$800\,\mathrm{GeV}$ energy fluxes and the best-fitting spectral energy distributions (SEDs) provided by \citet{ajello2020}. We also include an estimate of the Milky Way (MW) $\gamma$-ray luminosity, computed from a CR propagation model \citep{strong2010,Ackermann2012}. 
A power-law fit to these data as a function of the SFR, $L_{\mathrm{\gamma}}=A \dot{M}_*^m$, yields a value of $m = 1.42 \pm 0.17$ and $\log{A} = 39.21 \pm 0.15$, consistent with those reported in K20.

\subsection{The $L_{\mathrm{1.4 \, GHz}}$--SFR relation} \label{sec:L1.4_vs_SFR}

To obtain a radio-continuum luminosity sample centred at 1.4~GHz ($L_{\mathrm{1.4 GHz}}$) as homogeneous as possible, we measured the radio fluxes directly from available images whenever possible.
For eight galaxies (\object{NGC 253}, \object{NGC 1068}, \object{NGC~2146}, \object{NGC 2403}, \object{M82}, \object{Arp220}, \object{Arp299}, \object{NGC 3424}) continuum images at 1.4~GHz of Stokes-I are available in the NRAO/VLA Sky continuum Survey catalogue \citep[NVSS,][]{Condon1998}. For all the radio images, with the exception of \object{NGC 253}, we obtained an average rms ($1\sigma $) background fluctuation value of $\sim$ 0.5 mJy $\rm beam^{-1}$, consistent with what was reported by \citet{Condon1998}. We obtained the fluxes at 1.4 GHz ($F_{\mathrm{1.4 \, GHz}}$) using the radio interferometry data reduction package MIRIAD \citep{MIRIAD}, integrating the emission above $3\sigma$ to include all the flux related to the star formation and not only their central core and nucleus. We estimated the flux errors by integrating the emission above $2\sigma$ ($F_{\mathrm{2\sigma}}$) and subtracting it from the total flux.
Whenever possible, we compared our fluxes with those reported by \citet{Yun2001}, although these authors do not report flux errors. Our values are consistent with their nominal values. 

For the particular case of \object{NGC 253}, we excluded two background sources not associated with its intrinsic emission \citep{Kapinska2017}. Due to the extension of \object{NGC 2403}, we calculated the radio flux within the  optical isophote of $25\ \mathrm{mag\ arcsec}^{-2}$. We excluded patchy emission clearly disconnected from the main source and not associated with the optical emission of the galaxy, assuming that it does not come from the stellar populations.
For \object{NGC 4945} and \object{Circinus}, we calculate the 1.4~GHz flux using the same method but with the new Stokes-I continuum image provided by The Rapid Square Kilometre Array Pathfinder (ASKAP)  Continuum Survey\footnote{https://research.csiro.au/casda/} \citep[RACS,][]{McConnell2020}. For both galaxies we obtain an rms background of $\sim 0.3$~mJy/beam, consistent with the mean value $\sim 0.25$~mJy/beam reported by \citet{McConnell2020}. The nearby galaxy \object{Circinus} shows extended radio lobes \citep{Wilson2011}; we report an average flux between the one including the lobes (1.94 Jy) and the flux excluding them (1.51 Jy) in order to be consistent with the calculated fluxes for unresolved galaxies. We note that, in these cases, the fluxes we calculate may be slightly contaminated by the emission produced by AGN jets or other outflows that may arise in SBGs.

For the closest galaxies, \object{M31}, \object{M33}, and the Magellanic Clouds (SMC and LMC), the flux calculation is more complex due to the large spatial extent of these galaxies and the presence of resolved structures. Therefore, we adopt the fluxes provided by detailed studies of their emission available in the literature \citep{Dennison1975, Loiseau1987, Hughes2007}. 
 
\begin{figure*}
    \centering
    \includegraphics[width=0.41\textwidth]{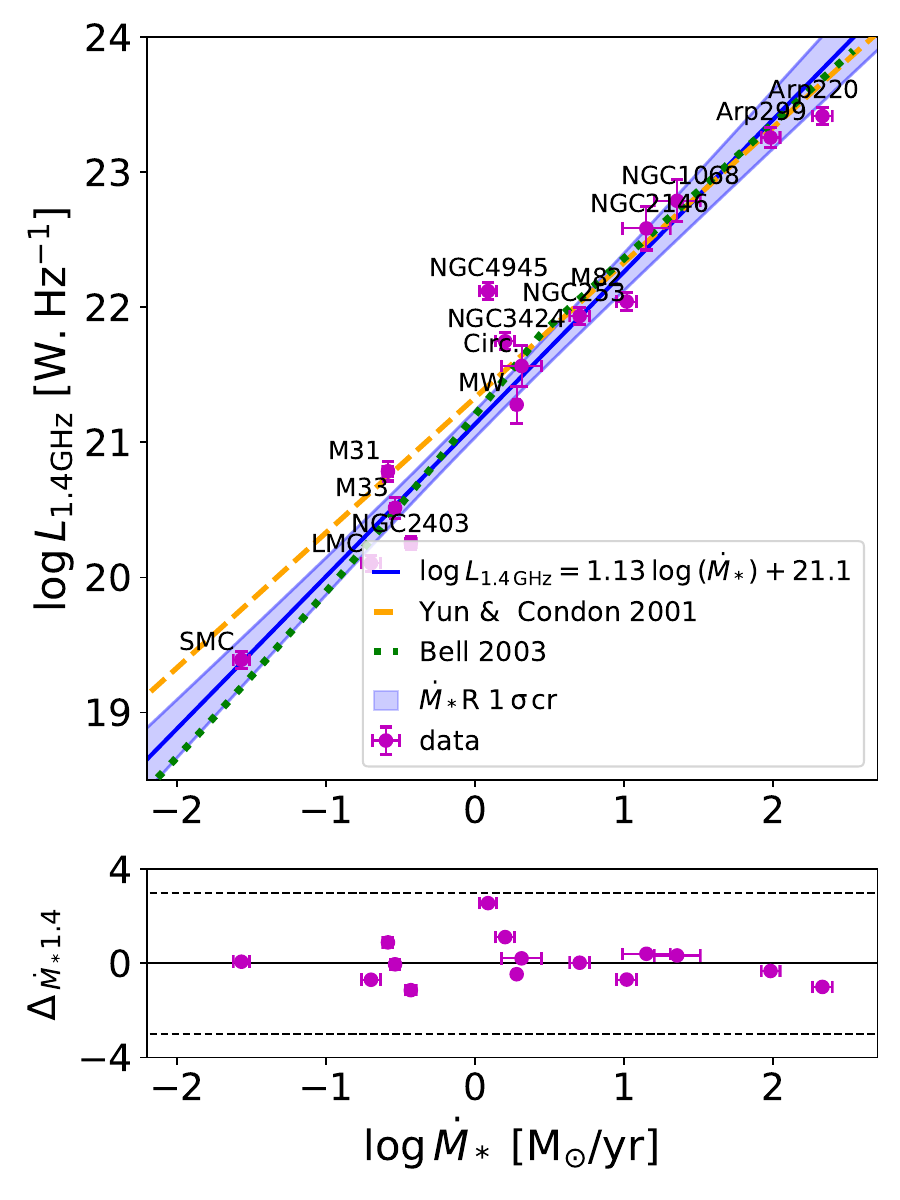}
    \includegraphics[width=0.41\textwidth]{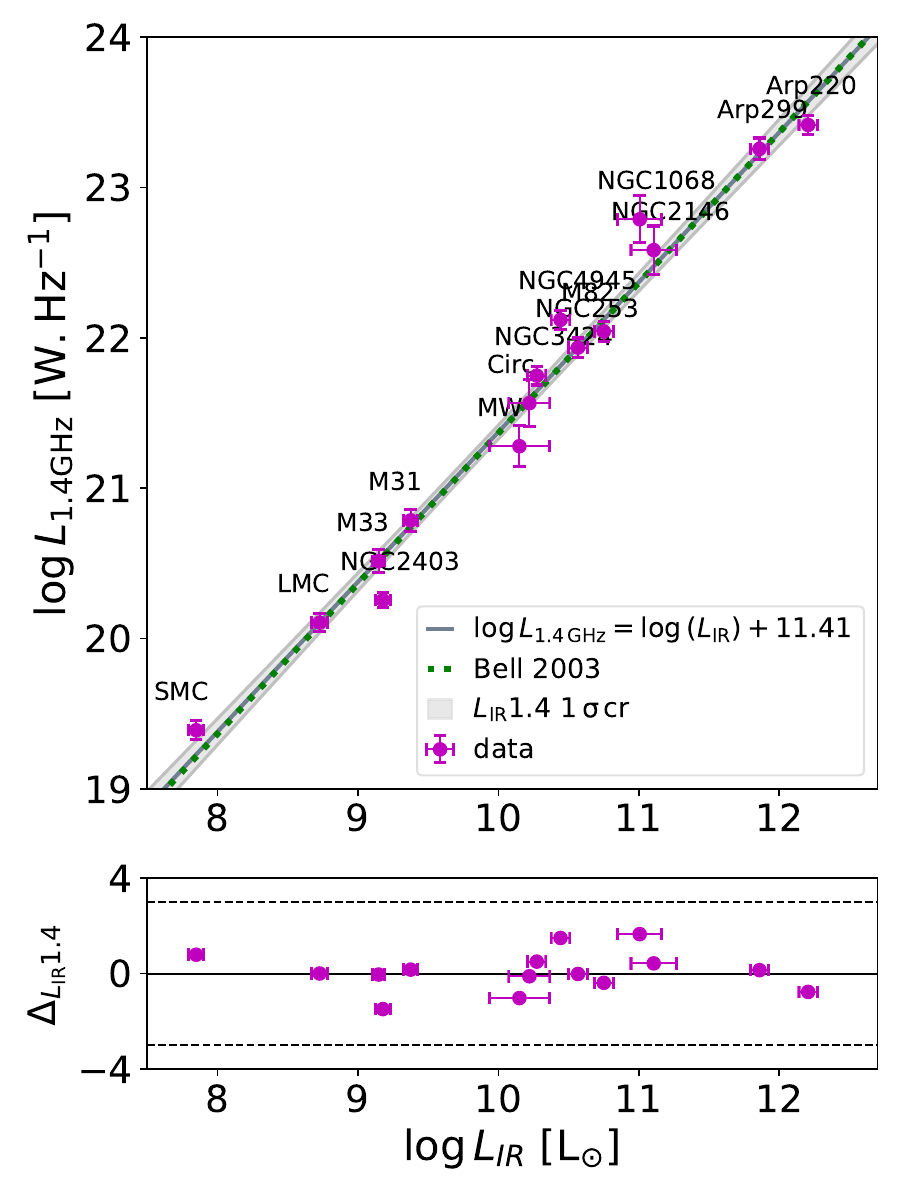}\\
    \includegraphics[width=0.41\textwidth]{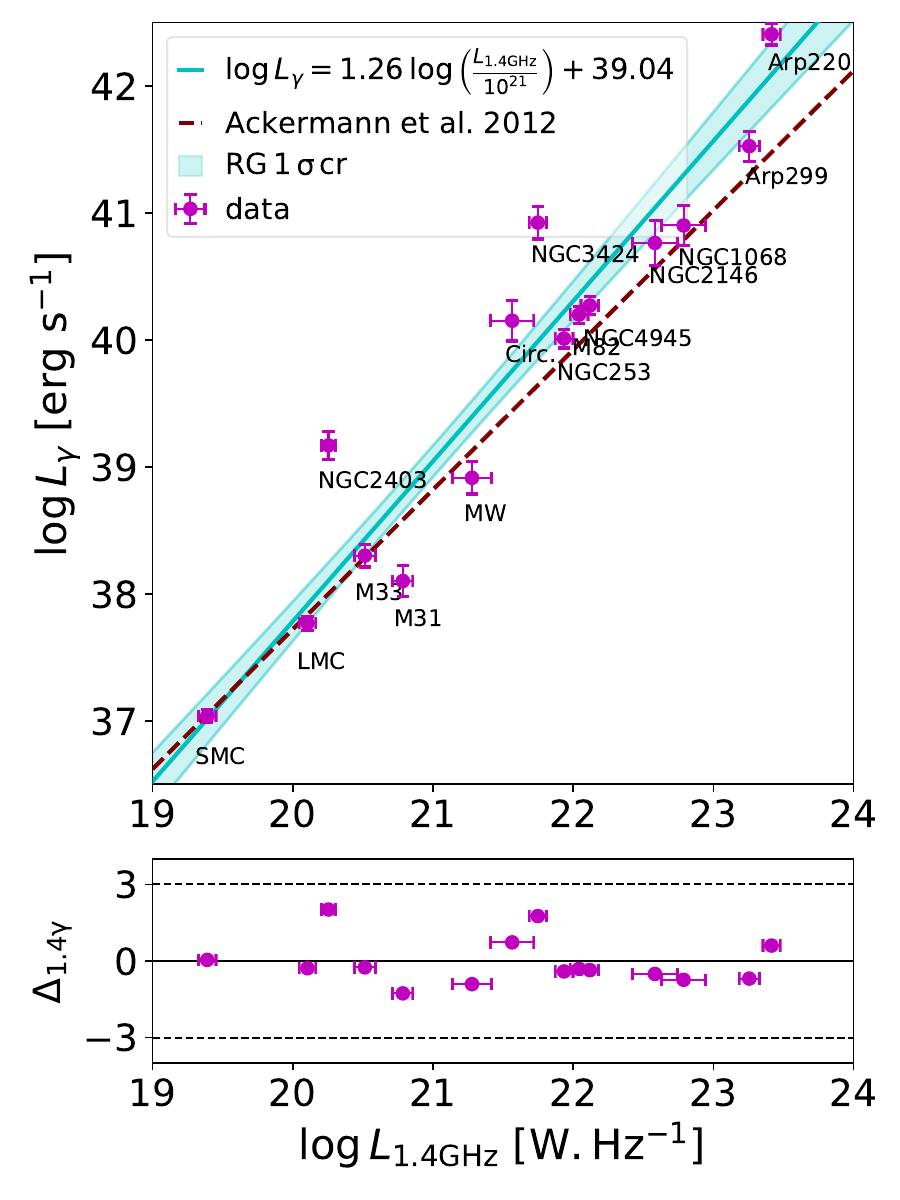}
    \includegraphics[width=0.41\textwidth]{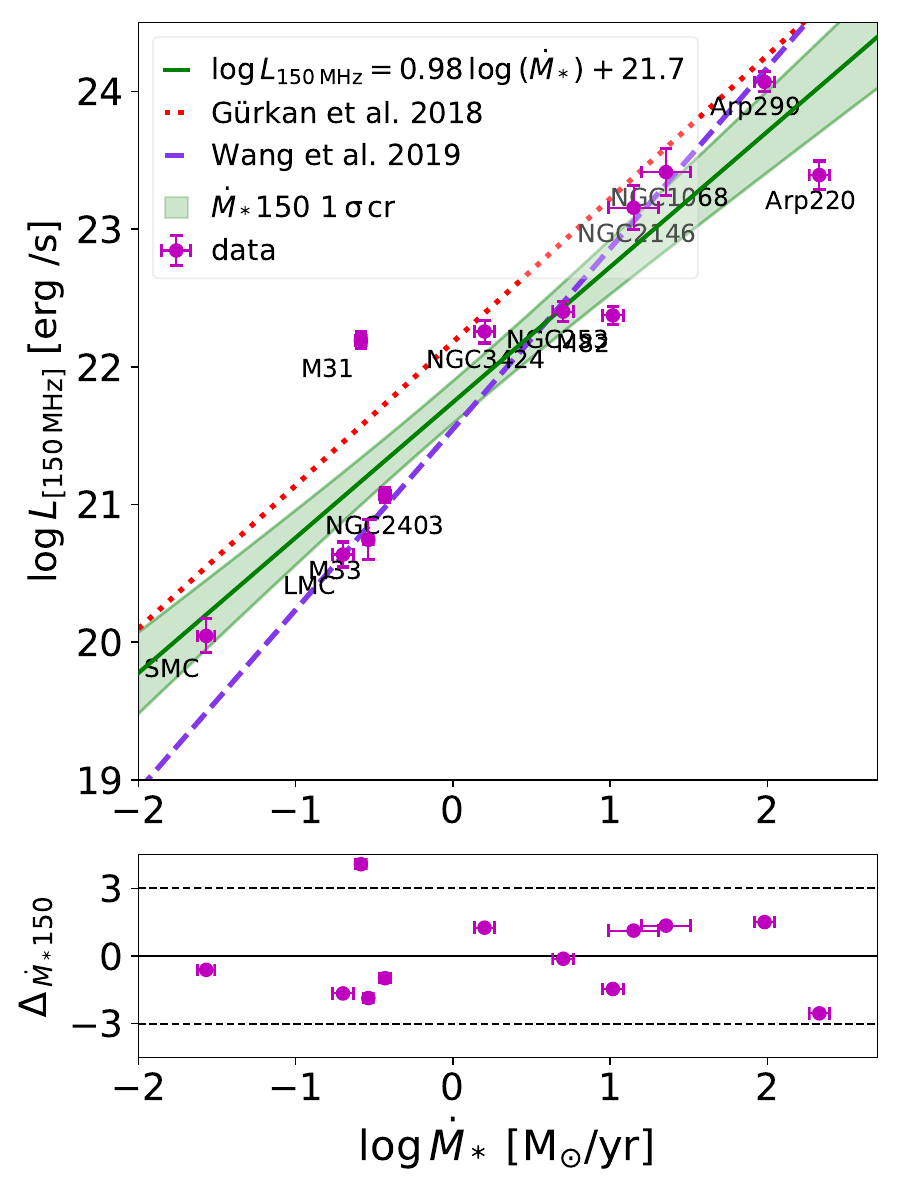}
    \caption{\textit{Upper left panel}: $L_{\mathrm{1.4 GHz}}$ as a function of the SFR. The blue solid line is the best fit to the data, and the shaded region its 68$\%$ confidence level. For comparison, the yellow dashed and green dotted lines show the \citet{Yun2001} and \citet{Bell2003} relations, respectively. \textit{Upper right panel:}  $L_{\mathrm{1.4 GHz}}$ as a function of $L_{\mathrm{IR}}$. The grey solid line is the best fit to the data, and the shaded region its 68$\%$ confidence level. For comparison, the green dotted line shows the \citet{Bell2003} relation. \textit{Lower left panel:} $L_{\gamma}$ as a function of $L_{\mathrm{1.4 GHz}}$. The cyan solid line shows the best fit to the data, and the shaded region its 68$\%$ confidence level. For comparison, the brown dashed line shows the \citet{Ackermann2012} relation \textit{Lower right panel:} $L_{\mathrm{150 MHz}}$ as a function of the SFR. The green solid line is the best fit to the data, and the shaded region its 68$\%$ confidence level. For comparison, the red dotted and violet dashed lines show the \citet{Gurkan2018} and \citep{Wang2019} relations, respectively. In all cases, the corresponding residuals weighted by their standard deviation are shown in the bottom panel; the black dotted lines correspond to
    3 standard deviations.}
    \label{fig: corr}
\end{figure*}

The total IR luminosity in the 8--1000~$\mu\mathrm{m}$ band ($L_{\mathrm{IR}}$) is usually used as a proxy for the SFR. A power-law fit to $L_{\mathrm{1.4 GHz}}$ as a function of $L_{\mathrm{IR}}$, $L_{\mathrm{1.4 GHz}}=A L_\mathrm{IR}^m$, with $L_{\mathrm{IR}}$ from K20, yields $m = 1.00 \pm 0.04$, $\log{A} = 11.41 \pm 0.37$ and a dispersion of 0.2~dex, indicating a very tight relation. These values are in excellent agreement with the the ones reported by \citet{Bell2003} for their sample of 162 SFGs, and are also consistent with the best-fit values for a much bigger sample of 1809 galaxies reported by \citet{Yun2001}. A power-law fit to SFR data, $L_{\mathrm{1.4 GHz}}=A \dot{M}_*^m$, yields a value of $m =1.13 \pm 0.09$ and $\log{A} = 21.14 \pm 0.10$, with a larger dispersion of 0.34~dex concentrated close to $\log{\dot{M_*}} \sim 0$. 

In the upper left panel of Fig.~\ref{fig: corr} we show the $L_\mathrm{1.4 GHz}$--SFR correlation derived from the galaxies in our sample. 
The galaxies follow a tight correlation except for \object{NGC 4945} which deviates from the relation marginally ($\Delta_{\dot{M_*}1.4}$ = 2.8). The latter may be due to an additional contribution linked to the hosted Seyfert 2 nucleus \citep{Lenc2009}, or a misguided estimate of the SFR (see discussion in K20). For comparison we also show the estimated correlation by \citet{Yun2001}. These authors derived their correlation estimating the SFR using the linear $L_\mathrm{IR}$--SFR relationship from \citet{Kennicutt1998}, which is not a good proxy for galaxies with low SFRs \citep{Bell2003}. This makes the $L_\mathrm{1.4 GHz}$--SFR relationship presented by \citet{Yun2001} inconsistent with our fit in the low SFR range. We show the latter on upper right panel of Fig.~\ref{fig: corr}.

The previous analysis shows that the luminosities we calculate for the restricted sample of galaxies detected in $\gamma$-rays are consistent with those obtained by other authors, and that they follow the global trend found for more general samples of SFGs observed in radio. This supports that the global radio emission of these galaxies does not present any particular characteristic with respect to that shown by more general samples of galaxies.

As a by product, a fit of a power law to the relation between $\gamma$-ray and radio luminosities ($L_{\gamma}=A(L_{\mathrm{1.4 GHz}}/{10}^{21}\,\mathrm{W\,Hz^{-1}})^m$) gives $m =1.26 \pm 0.10$, $\log{A} = 39.04 \pm 0.10$ and a dispersion of 0.5~dex. This slope is marginally steeper than the value previously reported by \citet{Ackermann2012}, who finds a slope of 1.10 $\pm$ 0.05. These differences may be due to the size of our sample, which is almost twice as large as that of \citet{Ackermann2012}, and also to the self-consistent and updated values of distances and fluxes we use. We show the resulting correlation on the lower left panel of Fig.~\ref{fig: corr}, together with the \citet{Ackermann2012} relation.

\subsection{The $L_{\mathrm{150 MHz}}$--SFR}\label{sec:L150_vs_SFR}

At low radio frequencies ($\lesssim$ 300 MHz) some SFGs show a flattening of their spectra. This may be caused by partial absorption of the synchrotron emission by ionised gas, or by energy losses and propagation effects of CRs \citep{Marvil2015, chyzy2018}. Therefore, low-frequency data can be useful to infer the absorbing mechanisms and properties of the medium.

Eleven galaxies in our sample (\object{M31}, \object{NGC 253}, \object{SMC}, \object{M33}, \object{LMC}, \object{NGC 2146}, \object{NGC 2403}, \object{M82}, \object{NGC 3424}, \object{Arp 299}, \object{Arp 220}) have fluxes at 150~MHz available in the literature. For NGC~1068 we take the flux at 145~MHz and the spectral index given by \citet{Kuehr1981}, and extrapolate the flux to 150~MHz. In all cases, we derive the corresponding luminosities at this frequency ($L_\mathrm{150\, MHz}$) using the K20 distances. We report these values in Table~\ref{tabledata}, together with the respective references. In the cases of M31 and Arp~299, we assumed a typical flux error of 10\%.

A power-law fit to the data, $L_{\mathrm{150 MHz}}=A \dot{M}_*^m$, yields a value of $m =  0.98 \pm 0.13$ and $\log{A} = 21.7 \pm 0.15$, with a dispersion of 0.5~dex. On the lower right panel of Fig.~\ref{fig: corr} we show the derived $L_\mathrm{150 MHz}$--SFR correlation, together with those estimated by \citet{Wang2019} and \citet{Gurkan2018}. Different works studied this correlation using much larger samples \citep{calistro2017,Read2018, Smith2020}, reporting values are between $m = 1.07$ \citep{Gurkan2018} and 1.37 \citep{Wang2019}. Our slope is compatible with that of \citet{Gurkan2018} but our data points fall systematically below the fit of these authors. Given the precision of radio data, this points to a systematic difference in the SFRs whose origin may be the different estimators used (UV--IR proxies in our case, multiwavelength photometric fits in theirs). Indeed these authors state that the SFRs provided by their method differ in $\sim 0.2$~dex from those obtained from H$\alpha$ luminosities, which are in general agreement with ours (see K20). A correction of 0.2~dex in SFR to their fit would bring it into agreement with our data, strongly suggesting that the difference in SFR estimation methods is responsible for this systematic departure.
Regarding \citet{Wang2019}, the value that they report is derived from a sample using only galaxies above their completeness limit ($\dot{M}_* \approx 3\, \mathrm{M}_\odot\, \mathrm{yr}^{-1}$). Their best fit using all the data, which has better statistics due to the larger number of galaxies at low SFRs, has a lower slope of $m \approx 1.1$ in better agreement with ours.
Although the 150~MHz data are not homogeneous because the fluxes were obtained from different surveys, and with different criteria for the flux density integration, they deserve to be studied because they provide unique constraints to the properties of non-thermal emission of SFGs. We will keep in mind this caveat when interpreting results based on these data.

%
\section{The model} \label{Sec: model}

%

We develop a non-thermal emission model to understand the observed $L_{1.4\,\mathrm{GHz}}$--SFR  correlation (Sec.~\ref{sec:L1.4_vs_SFR}), while also contemplating the $L_\gamma$--SFR (Sec.~\ref{sec:Lgamma_vs_SFR}) and $L_{\mathrm{150\,MHz}}$--SFR (Sec.~\ref{sec:L150_vs_SFR}) relations.
The fact that two integrated properties of galaxies are involved in these correlations ---the integrated non-thermal luminosity and the SFR--- makes it reasonable to describe their emission based on the global CR energy balance. 
It is sensible to address the comprehension of the correlations between the non-thermal luminosity and the SFR in terms of a population of non-thermal particles that is injected at a rate that depends on the SFR of the galaxy. We further assume that the galactic environment in which these particles are injected is homogeneous and stationary. We self-consistently calculate the energy losses, diffusion and advection of these relativistic particles and their non-thermal radiative output, corrected by the relevant absorption effects. In what follows we detail how we incorporate all this physics in a model valid over a broad range of SFRs.

\subsection{CR injection and transport}
We model the galaxy emission region as a disk of radius $R$ and height $2H$ with constant gas density, and in which acceleration sites are uniformly distributed. We adopt $R = 1\, \mathrm{kpc}$ and $H=0.2\, R$ suggested by K20.
Following the standard prescriptions of diffusive shock acceleration at SNRs \citep[e.g.][]{Axford1977, Bell1978, Blandford1978}, we assume that CRs are steadily injected in the system according to:
\begin{equation}
 Q_i(E)=Q_{0,i} E^{-\alpha} \exp{\left[-(E/E_{\mathrm{max}, i})^{\zeta_i}\right]},
\end{equation}
where $i=\mathrm{e,p}$ (electrons and protons), $\alpha=2.2$ is the spectral index \citep{Lacki2013}, $E_\mathrm{max}$ is the maximum particle energy (assumed to be $10^{15}$~eV), and $\zeta_\mathrm{p,e} = (1,2)$ \citep{Zirakashvili2007, Blasi2010}.
The normalisation constant $Q_{0,i}$ is computed as
\begin{equation}
    \mathcal{R}_{\rm SN} E_{\rm SN} \xi_{\rm SN} = \int_{E_{\mathrm{min},p}}^{E_{\mathrm{max},p}}  E \,  Q_{0,p}(E) \,  dE,
    \label{energynorm}
\end{equation}
assuming that the supernova rate of the galaxy is $\mathcal{R}_{\mathrm{SN}}= (83 \, \mathrm{M}_{\odot})^{-1} \dot{M}_*$, consistent with the \citet{chabrier2003} initial mass function (IMF), and that a fraction $\xi_{\rm CR} = 10 \%$ of the supernova energy ($E_{\rm SN} = 10^{51} \, \rm erg$) is converted in CRs. 
We adopt $E_{\mathrm{min,p}}$ = 1.2 GeV as the minimum proton energy. 
Moreover, this energy is distributed as $Q_{0,p}/Q_{0,e} = 50$; such a proton-to-electron ratio is supported by multiwavelength modelling and observations \citep{Torres2004, Merten2017, Peretti2019, Yoast-Hull_M82}. We note that a slightly different value compared to our benchmark assumption (within a factor two), only affects the emission from primary electrons (also within a factor two), which does not impact significantly our global results.

Young massive star clusters, as discussed in \cite{Morlino2021} can be additional acceleration sites for CRs where diffusive shock acceleration is inferred to be efficient possibly up to PeV. 
Even though these sources can play an important role as acceleration sites, it is not clear yet to which extent they can contribute to the bulk of galactic CRs where, so far, SNRs are believed to be dominant. 
Therefore, we work under the assumption that SNRs are dominating the injection of the bulk of CRs. 
Nevertheless, a second population of galactic accelerators such as young massive star clusters is unlikely to change our main results as far as the acceleration efficiency of both components is adjusted to produce a total CR power similar to that used here.

Electrons and protons experience different losses, which can be divided into radiative, non-radiative, and escape.
Electrons suffer radiative losses via synchrotron, inverse Compton (IC) up-scattering of IR photons, and relativistic Bremsstrahlung (BS). 
Protons lose energy via inelastic $p$-$p$ collisions.
For both particle species the non-radiative losses are due to ionisation. 
We refer to K20 for the exact expressions used to calculate the corresponding timescales. In particular, for the IC process we adopt the formalism developed by \cite{Khangulyan2014} which accounts for the complete differential cross-section in both Thomson and Klein -- Nishina regimes.
Finally, diffusion and advection regulate the particle escape from the emission region. We discuss the details of escape timescales in the following section.

The balance between injection and loss mechanisms leads to the following approximate steady-state solution for the particles distribution in the galaxy:
\begin{equation}
 N_i(E) = \frac{1}{\mid \dot{E} \mid} \int_E^{E_{\mathrm{max}, i}} Q_i(E) \, dE ,
 \label{Eq: Transport}
\end{equation}
where $\dot{E} = E\, \tau_\mathrm{loss}^{-1}(E)$ and
$\tau_\mathrm{loss}~=~1/[\tau_\mathrm{esc}^{-1} + \tau_\mathrm{cool}^{-1}]$. Here $\tau_\mathrm{esc}$ is the characteristic CRs escape time computed as $\tau_\mathrm{esc}^{-1} = \tau_\mathrm{adv}^{-1} + \tau_\mathrm{diff}^{-1}$ and $
\tau_\mathrm{cool}^{-1} = \sum_j \tau_j^{-1}$, where each $\tau_j$ represents the CRs cooling time by each different cooling mechanisms.

We note that this equation is almost equivalent to the standard leaky box solution, $N_i(E) = Q_i(E) \, \tau_{\rm loss}(E)$, with the benefit that it yields a more precise normalisation for injection indices $\alpha > 2$ when cooling dominates over escape.

\subsection{Secondary products}
\label{SubsecSec: Seconday}

Here we present the formulae used to calculate $\gamma$ rays and secondary electrons spectra produced in $p$-$p$ interactions. 
Following \citet{Kelner2006}, the emissivity of these secondary products $i=\gamma$,e can be summarised in two different formalisms depending on the energy $E_i$ of the resulting particle. In the following equations we denote $x_i=E_i/E_p$. 

For $E_{\mathrm{i}} > 100$~GeV the emissivity formula is:
\begin{equation}
Q_i(E_i)=c \, n \int^1_{x_{i, \mathrm{min}}} \sigma_\mathrm{pp}\left(\frac{E_i}{x_i}\right)\,N_p\left(\frac{E_i}{x_i}\right)\,\tilde{F}_i\left(x_i,\frac{E_i}{x_i}\right)\, \frac{dx_i}{x_i},
\end{equation}
with $\sigma_\mathrm{pp}$ the cross section for inelastic $p$-$p$ scattering,
\begin{equation}
\sigma_\mathrm{pp}(E) = \left(34.3 + 1.88 L + 0.25 L^2\right) \times \left[1-\left(\frac{E_\mathrm{th}}{E} \right)^4 \right]^2 \,\mathrm{mb},
\end{equation}
where $E_\mathrm{th} = 1.22$~GeV is the threshold energy for $\pi_0$ production, $L = \ln{(E/\mathrm{TeV})}$ and $\kappa = 0.5 $ is the process inelasticity.
$\tilde{F}_{\gamma}$ and $\tilde{F}_{e}$ are given in equations (58--61) and (62--65) of \citet{Kelner2006}, respectively. 

For energies $E_{\mathrm{i}} \leq 100$~GeV we describe the emissivity by the $\delta$-functional approximation \citep{Aharonian2000}:
\begin{equation}
\begin{aligned}
Q_i(E_i)=c\,n\int^\infty_{E_{i, \mathrm{min}}}\frac{2}{K_{\pi}} \sigma_\mathrm{pp}\left(\frac{E_{\pi}}{K_{\pi}}+m_p c^2\right) \, N_p\left(\frac{E_{\pi}}{K_{\pi}}+m_p c^2\right)\\ \times\frac{\tilde{f_i}(E_i/E_{\pi})}{\sqrt{E_{\pi}^2-m_{\pi}c^2}}\, dE_{\pi}.
\end{aligned}
\end{equation}
Here $n$ is the number density of the ISM, $m_p$ and $m_\pi$ are the masses of the proton and the pion, respectively, $E_{\pi}$ is the pion energy, $E_{i,\mathrm{min}} = E_i + m_{\pi}^2 c^4 /4E_i$, $K_{\pi} \approx 0.17$ is the fraction of kinetic energy transferred from the parent proton to the single pion, $\tilde{f_{\gamma}}(E_{\gamma}/E_{\pi}) =1$ 
and $\tilde{f_{e}}(E_e/E_{\pi})$ are defined in equations (36--39) from \citet{Kelner2006} (see also Appendix~\ref{Appendix-Kelner}). 

\subsection{Free-free absorption and emission}

\label{subsec: free-free}
 
Part of the ionised gas in SFGs fills uniformly the ISM, whereas another part is concentrated in H{\sc ii} regions randomly distributed over the disc. This gas emits free-free radiation in the radio band. To estimate this emission, we consider that the ionised gas is homogeneously distributed in the same volume as the CRs.
 In addition, this gas can be optically thick to the low-frequency (thermal and non-thermal) radio emission. Our model has two free parameters that determine the impact of the free-free absorption and emission processes: the electron temperature $T_e$ and the ionisation fraction $\rho$, defined here as the ratio between ionised and total number densities.
Thus, we compute the absorbed free-free thermal luminosity at a frequency $\nu$, $L_\mathrm{ff}(\nu)$, following the \citet{Olnon1975} formalism for an homogeneous cylinder seen face-on, valid in the Rayleigh-Jeans regime ($h \nu \ll k T_e$, where $h$ and $k$ are the Planck and Boltzmann constants, respectively),
\begin{equation}
\label{eq: thermalff}
L_\mathrm{ff}(\nu) = \frac{8 \pi^2 k T_{\mathrm{e}}\nu^2}{c^2}R^2 \left(1 - e^{-\tau_\mathrm{ff}(\nu)} \right),
\end{equation}
where $\tau_\mathrm{ff} = \alpha_\mathrm{ff}(\nu)\,l$ is the 
optical depth, $l = 2H$ is the length of the ionised region along the line of sight and $\alpha_{\mathrm{ff}}$ is the absorption coefficient for photons of frequency $\nu$. 
This coefficient is given by \citet{Rybicki1986},
\begin{equation}
\label{eq: absorption}
\alpha_{\mathrm{ff}} = 3.7 \times 10^8 \rho^2 n^2 T_{e}^{-1/2} Z^2 \nu^{-3} \left[1-\exp{\left(-\frac{h\nu}{kT_{e}}\right)} \right]\bar{g}_\mathrm{ff},
\end{equation}
where $T_{\mathrm{e}}$ the ionised gas temperature and $\bar{g}_\mathrm{ff}$ is the mean Gaunt factor  \citep[e.g.][]{Leitherer1995},
 \begin{equation}
 \bar{g}_\mathrm{ff}=9.77 \left(1 + 0.13 \, \mathrm{log} \frac{T_{e}^{3/2}}{Z \nu}\right)
 \end{equation}
 
\noindent
Following \citet{ISM_book}, we assume a constant ionisation fraction $\rho= 0.05$ throughout the whole SFR range, and we explore the impact of higher values, up to 0.1, on the luminosity--SFR correlations. 
In SBGs and ULIRGs the ISM conditions are less well-known. Despite the high star forming activity and supernova rate, most of the ionised component may be removed from the central galactic region and load galactic winds. 
This would lead to extremely low values of the ionisation ratio down to $10^{-4}$ \citep[][]{Krumholz2020}. 
We take this into account exploring scenarios where $\rho$ is assumed to such a low value.

The warm neutral and ionised medium phases of the ISM are inferred to have an average temperature of $10^4$ K. However, depending on the fraction of cooler neutral or hot ionised component, the effective ISM temperature could possibly differ by about one order of magnitude \citep{Quireza2006}. Therefore, we assume $T_{\mathrm{e}} = 10^4$ K as a reference value and explore the impact of an order of magnitude change in its value.

In a similar manner, we obtain the outgoing non-thermal flux of the whole emission region by solving the corresponding radiative transport equation \citep[see][]{Rybicki1986}.
Modelling the emission region as a homogeneous face-on cylinder, the absorption-corrected synchrotron luminosity is:
\begin{equation}
L_\mathrm{abs}(\nu) = L_\mathrm{0}(\nu)\left( \frac{1 -e^{-\tau_\mathrm{ff}(\nu)}}{\tau_\mathrm{ff}(\nu)} \right),
\end{equation}
\noindent where $L_{0}$ is the unabsorbed luminosity produced by non-thermal radiative processes in our model. 

 %
 
%
\section{Modelling and scale relations}
\label{Sec: Model and Scale relations}
%
The SFR is the only independent variable in our model. We assume that the energy injected into protons and electrons depends linearly on the SFR (Eq.~\ref{energynorm}). 
The observed correlations presented in Sec.~\ref{Sec: obs_correl} differ from linearity, which suggests that there is also a dependence of the main galactic properties on the SFR.
In this section, we are interested in finding simple parametrisations for the density $n$, the IR radiation field, and the magnetic and velocity fields with the SFR, motivated by physical and observational arguments. 

\subsection{Densities and $L_\mathrm{IR}$}
We link the IR luminosity to the SFR by using the relation of \citet{Bell2003}, adapted to a \citet{chabrier2003} IMF:
\begin{equation}
\dot{M_*}[\mathrm{M}_\odot\,\mathrm{yr}^{-1}]=1.72 \times 10^{-10} \frac{L_\mathrm{IR}}{L_\odot} \left(1 + \sqrt{\frac{10^{9}}{L_\mathrm{IR}/L_\odot}}\,\right)\epsilon ,
\end{equation}
with $\epsilon=0.79$ \citep{crain2010} and $L_\mathrm{IR}$ the total infrared luminosity between 8 and 1000~$\mu$m.
For the calculation of the IC interactions we introduce a dilution factor of the energy density of the IR radiation field with respect to the one corresponding to a black body. For a disc geometry this factor is $\Sigma = L_{\rm IR}/(8 \pi R^2 \sigma_\mathrm{SB} T^4) $, where $\sigma_\mathrm{SB}$ is the Stephan-Boltzmann constant.

Finally, the scaling of the density $n$ with the SFR can be set by using the Kennicutt-Schmidt (K-S) law \citep[][]{Kennicutt1998,2012ARA&A..50..531K,2019ApJ...872...16D}. This law gives the correlation between the surface densities of SFR ($\Sigma_\mathrm{SFR}$) and cold gas mass ($\Sigma_\mathrm{gas}$) of galaxies. Assuming that the emitter is a cylinder of radius $R$ and thickness $2H$ (K20), we get
\begin{equation}
    \label{eq: dens_sfr}
    n = \frac{\Sigma_\mathrm{gas}}{2 H m_\mathrm{H}} \propto \dot{M}_*^{0.71}\,R^{-1.42}\,H^{-1}.
\end{equation}

\subsection{Magnetic field}

\citet{Robinshaw-2008} showed that strong magnetic fields ($B$ of the order of mG -- hundreds times stronger than in our Galaxy) pervade the ISM of high SFR objects, such as ULIRGs, suggesting a strong dependence of $B$ on the level of source activity.
Therefore, a detailed modelling of the transport of CR-electrons and their associated non-thermal radio emission requires a careful treatment of the magnetic field and its dependence on the SFR.
For this reason, we improved the model developed in K20 proposing the following scaling,
\begin{equation}
    \label{Eq: Magnetic-field}
    B = B_0 \, \left( \frac{\dot{M}_*}{10 \, \rm M_{\odot} \, {\rm yr}^{-1} } \right)^{\beta}.
\end{equation}
where the normalisation factor $B_0$ is a free parameter and for the index $\beta$ we explore three possible scenarios, $\beta = 0.3, 0.5, 0.7$, based on equipartition arguments described below.

The weakest scaling is obtained relying on a stationary condition of the MHD turbulence in the ISM plasma.
The MHD turbulence reaches its stationary state when the magnetic field growth produced by turbulence-induced field line stretching is balanced by the back-reaction of the magnetic field itself \citep{Groves2003}. 
Such a condition results in equipartition between the energy density of the magnetic field and the kinetic energy density of the gas. 
This yields a scaling $B \propto \Sigma_{\rm g}^{0.5}$, which is in good agreement with numerical simulations that report exponent values in the range 0.4--0.6 \citep{Groves2003}.
By applying the K-S scaling ($\Sigma_\mathrm{SFR} \propto \Sigma_\mathrm{g}^{1.41}$) one can retrieve the index $\beta \approx 0.3$.
The second scenario is based on an equipartition condition between magnetic field and starlight, namely $U_{\rm B} \approx U_{\rm rad}$. 
This condition, often quoted in the literature \citep[e.g.][]{Murphy2009,Lacki2010,Yoast2016}, is motivated by the assumption that the radiation pressure drives the turbulence in the ISM, which in turn determines the strength of the magnetic field. The linear relation between the SFR and the starlight results in the scaling $\beta \approx 0.5$.
The third scenario considered relies on the stability of the galactic disc. As discussed by \citet{Lacki2010} \citep[see also][]{Parker1966}, the equipartition between the gas pressure at the mid-plane of the disc ($\pi G \Sigma_g^2 $) and the magnetic field pressure ($B^2/8\pi$) provides a natural limit for the magnetic field strength; higher values would make the galactic disk unstable. 
This condition, once the Kennicutt scaling is applied, leads to $\beta=0.7$.

We assumed $\beta=0.3$ as the fiducial value for the magnetic field index. In addition, we explored other prescriptions with a stronger $B$ scaling ($\beta=0.5, 0.7$), which could be consistent with a more relevant molecular phase of the ISM in environments with high SFR. 
We adopted $B_0=100~\mu{\rm G}$ as a fiducial normalisation, which yields $B$-values in agreement with both the expected ones for normal starbursts galaxies \citep[see also][]{Peretti2019} and with the one inferred for the Galactic central region \citep[see][]{Ferriere-2009, Lacki2013}. We also explored different values of $B_{0}$ across an order of magnitude to cover the different inferred values for the average magnetic field in SFGs, ranging from $\sim \mu$G in low SFR galaxies \citep{Ferriere2010,Jurusik2014} to $\sim $mG for ULIRGs \citep{Robinshaw-2008}.

\subsection{Advection}

Winds are ubiquitous in SFGs and are possibly one of the most spectacular consequences of their star forming activity. 
They are capable of driving mass out of galaxies at a rate of several $\rm M_{\odot} \, yr^{-1}$ with velocities up to $10^3 \, \rm km \, s^{-1}$ \citep[see][]{Veilleux_2005}. 
The mechanism(s) powering galactic winds are likely to depend on the SFR. In SBGs they arise from the combined effect of mechanical energy and heating produced by supernovae, and radiation pressure from young stars \citep{Zhang_2018}. On the other hand, these effects are unlikely to be the main responsible for wind launching in galaxies with mild SFR; instead, the CR pressure seems to play the key role. In particular, the gradient of escaping CRs can be strong enough to overcome the gravitational pull and lift the ISM into the halo \citep{Breitschwerdt_1991,Recchia_2016}. In addition, in the case of SBGs, CRs cool more efficiently via $p$-$p$ collisions and their contribution to powering galactic winds could be minor, but this is still under debate given the importance of the CR pressure exerted on the cold interstellar gas \citep{Bustard_Zweibel2020}.

The effect of galactic winds on the transport of CRs needs to be taken into account. We thus consider that particles are advected away from the galaxy with an energy-independent characteristic time:
\begin{equation}
    \label{Eq: Adv-timescale}
    \tau_{\rm adv}(\Dot{M}_*) = \frac{H}{v_{\rm adv}(\Dot{M}_*)}
\end{equation}
where $v_{\rm adv}$ is the SFR-dependent advection speed.

We divide the whole galaxy population in two classes: 1) high SFR and 2) mild SFR objects. We assume that objects with a star formation higher than the MW, $\Dot{M}_* \gtrsim 2 \, \rm M_{\odot} \, yr^{-1}$, are characterised by a starburst-driven wind, $v_\mathrm{sw} = 400 \, \mathrm{km \, s^{-1}}$ \citep[see also K20,][]{Yoast-Hull_M82,Peretti2019}. 
For objects with $\Dot{M}_* < 2 \, \rm M_{\odot} \, yr^{-1}$, we assume that the galactic wind is CR-driven.  
In this scenario, \cite{Recchia_2016} have shown that, at the base of the wind, the dominant advection velocity is the Alfv\'en speed. Therefore, we consider that particles are advected away from the galactic box at the local Alfv\'en speed, $v_\mathrm{A}= B/\sqrt{4 \pi \, m_p \, n_i}$.
Putting all together, we define a global advection velocity over the entire SFR range as
\begin{equation}
    v_{\rm b}(\dot{M}) =
    \begin{cases}
    v_{\rm A} \,, & \Dot{M}_* < 2 \, \rm M_{\odot} \, yr^{-1} \\
    v_\mathrm{sw} \,, & \Dot{M}_* > 2 \, \rm M_{\odot} \, yr^{-1}
    \end{cases}
    .
    \label{eq:v_b}
\end{equation}

\noindent and we assume $v_{\rm adv} = v_{\rm b}(\dot{M}) $ to compute the Eq.~\ref{Eq: Adv-timescale}.
In addition, we explore alternative scenarios in which the nature of the advection speed is unique (either $v_\mathrm{A}$ or $v_\mathrm{sw}$) for the entire SFR range considered.

\subsection{Diffusion}

Diffusion is a fundamental aspect of CR physics. 
At a microscopic level, this phenomenon consists in resonant pitch-angle scattering of CRs with the local average magnetic field, taking place when the Larmor radius approaches the wavelength of local magnetic field disturbances \citep[e.g. Alfv\'en waves, as discussed in][]{Blasi-review}.
At a macroscopic level, this small scale interaction results in a diffusive motion of CRs, where the confinement properties of the system are parametrised by the diffusion coefficient, $D$.
This coefficient holds all information that effectively characterises diffusion, such as the turbulence properties of the medium.

We compute the diffusion coefficient adopting the quasi-linear theory formalism
 \begin{equation}
   D(E,\dot{M_*}) \approx \frac{1}{3} r_{L} \, v(E) \, \left( \frac{l_c}{r_L} \right)^{1-\delta},
   \label{Eq: diff-SFR}
\end{equation}
where $v(E)$ is the particle speed, $r_L$ is the Larmor radius, $l_c$ represents the coherence length of the magnetic field, and $\delta$ characterises the nature of the turbulence. 
In particular, we have $\delta=1/3$ for the Kolmogorov turbulence, and $\delta=1, \, 0.5$ for Bohm and Kraichnan. 
In this work we assume the Kolmogorov-like turbulence, $\delta=1/3$, as our benchmark case, in agreement with current observations of the boron-to-carbon ratio in the Galaxy \citep[][]{Aguillar2016}. We assume that supernovae are responsible for the injection of the turbulence in the system.
Therefore, we compute $l_c = (2H \pi R^2 \mathcal{R}_{\mathrm{SN}}^{-1}\tau_{\mathrm{SNR}}^{-1})^{1/3}$ as the average separation between SNRs in the galactic box, where we adopt a typical value for the SNR lifetime $\tau_{\mathrm{SNR}} = 3\times 10^{4} \, \mathrm{yr}$ \citep{Truelove1999}. 
We also impose that such a parameter cannot be smaller than the typical dimension of a middle age SNR, namely $l_c \gtrsim 10 \rm \, pc$.

The diffusion coefficient defined in Eq.~\eqref{Eq: diff-SFR} is assumed to be homogeneous in the whole galactic box.
This allows us to define the characteristic timescale, $\tau_{\rm D}$, in which particles diffuse away from the galaxy
\begin{equation}
    \label{Eq: Diff-timescale}
    \tau_{D}(E,\dot{M}_*) = \frac{H^2}{D(E,\dot{M}_*)}.
\end{equation}
We give the explicit dependency of $\tau_D$ on the SFR in Appendix~\ref{ap:model}.

\begin{table}
    \centering
    \begin{tabular}{lcccc}
    \hline \hline
    Scenario & $\beta$ & $B_0$ & $\rho$ & $T_e$[K]\\ \hline
    0 & 0.3 & 100 & 0.05  & $10^4$\\
    \hline
    1 & 0.5 & 100 & 0.05  &$10^4$\\   
    2 & 0.7 & 100 & 0.05  &$10^4$\\
    3 & 0.3 & 50 & 0.05   &$10^4$\\
    4 & 0.3 & 500 & 0.05  &$10^4$\\
    5 & 0.3 & 100 & $10^{-4}$ &$10^4$\\
    6 & 0.3 & 100 & 0.1   &$10^4$\\
    7 & 0.3 & 100 & 0.05  &$10^3$\\
    8 & 0.3 & 100 & 0.05  &$10^5$\\
    \hline \hline
    \end{tabular}
    \caption{Values of the free parameters of our model (index of $B(\dot{M}_*)$ relation, its normalisation, ionised gas fraction, temperature of the ionised gas) for all the scenarios considered.}
    \label{Table_modelparameters}
\end{table}

%
\section{Results and Discussion}
\label{Sec: Results}
To investigate how star formation influences the transport of CRs in SFGs, we set up a benchmark parameter configuration (scenario (0) in Table~\ref{Table_modelparameters}), to compute the global properties of galaxies throughout the whole SFR range ($10^{-2}$--$10^{3} \, \rm M_{\odot} \, yr^{-1}$), according to the scaling relations discussed in Sec.~\ref{Sec: Model and Scale relations}.
We compare the timescales of the different processes involved, and solve the transport equation (Eq.~\ref{Eq: Transport}) following the prescriptions of Secs.~\ref{Sec: model} and  \ref{Sec: Model and Scale relations}, to determine the main characteristics of CR transport in different ranges of SFR. We present the results for this scenario in Sec~\ref{subsec: General transport properties}. 

The transport of CRs shapes the non-thermal SED, and in turn each luminosity--SFR relation. We focus especially on the impact of different CR transport conditions on these relations.
We therefore explore the outcome of our model in terms of the photon SED, from which we extract the integrated luminosity in the \textit{Fermi-LAT} energy band (0.1--100 GeV) and the monochromatic luminosities at 1.4~GHz and 150~MHz. 

We define an upper bound to the luminosity  as a function of the SFR (calorimetric limit) in our model, to assess the ability of SFGs to cool CRs, converting their energy into radiation. 
Such a limit is built under the following conditions:
i) the escape is artificially turned off so that all injected particles cool inside the source;
ii) the non-thermal emission is not absorbed. 
This leads to maximum values $L_\gamma^\mathrm{max}$, $L_{1.4\mathrm{\, GHz}}^\mathrm{max}$ and $L_{150 \mathrm{MHz}}^\mathrm{max}$ for each SFR.

We finally analyse the robustness of the model by exploring its parameter space, in order to assess the impact of different assumptions on the luminosity--SFR correlations. 
The scenarios considered are listed in Table~\ref{Table_modelparameters}. 
Each scenario from (1) to (8) represents a single parameter variation with respect to the benchmark scenario.
We present, discuss and compare our results of each luminosity--SFR correlation in Secs.~\ref{subsec: result_gamma-sfr}, \ref{subsec: result_1.4-sfr} and \ref{subsec: result_150-sfr}, respectively.

\subsection{Transport properties of the benchmark scenario}
\label{subsec: General transport properties}

\begin{figure*}
    \centering
    \includegraphics[angle=-90, width=0.49\textwidth]{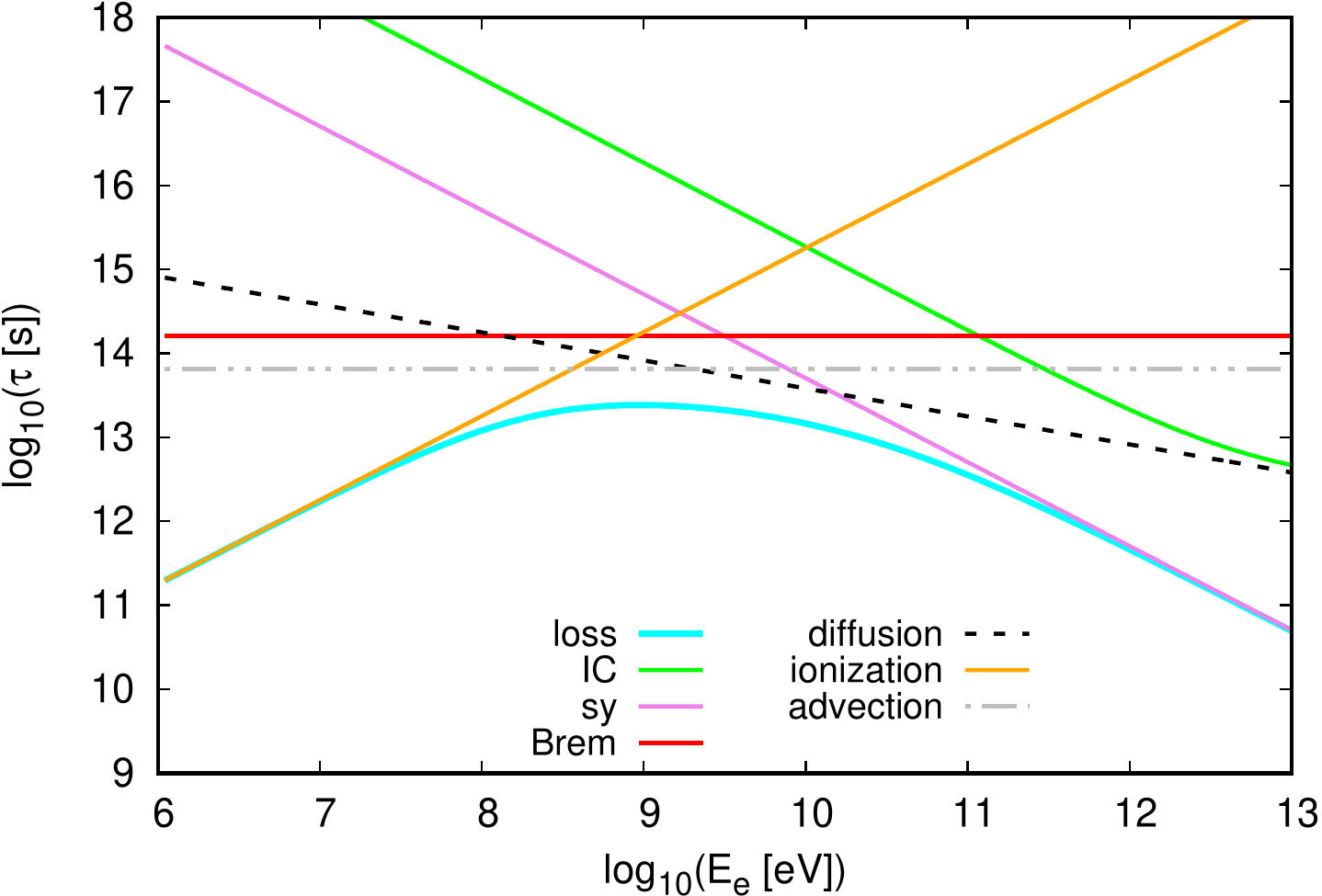}
    \includegraphics[angle=-90, width=0.49\textwidth]{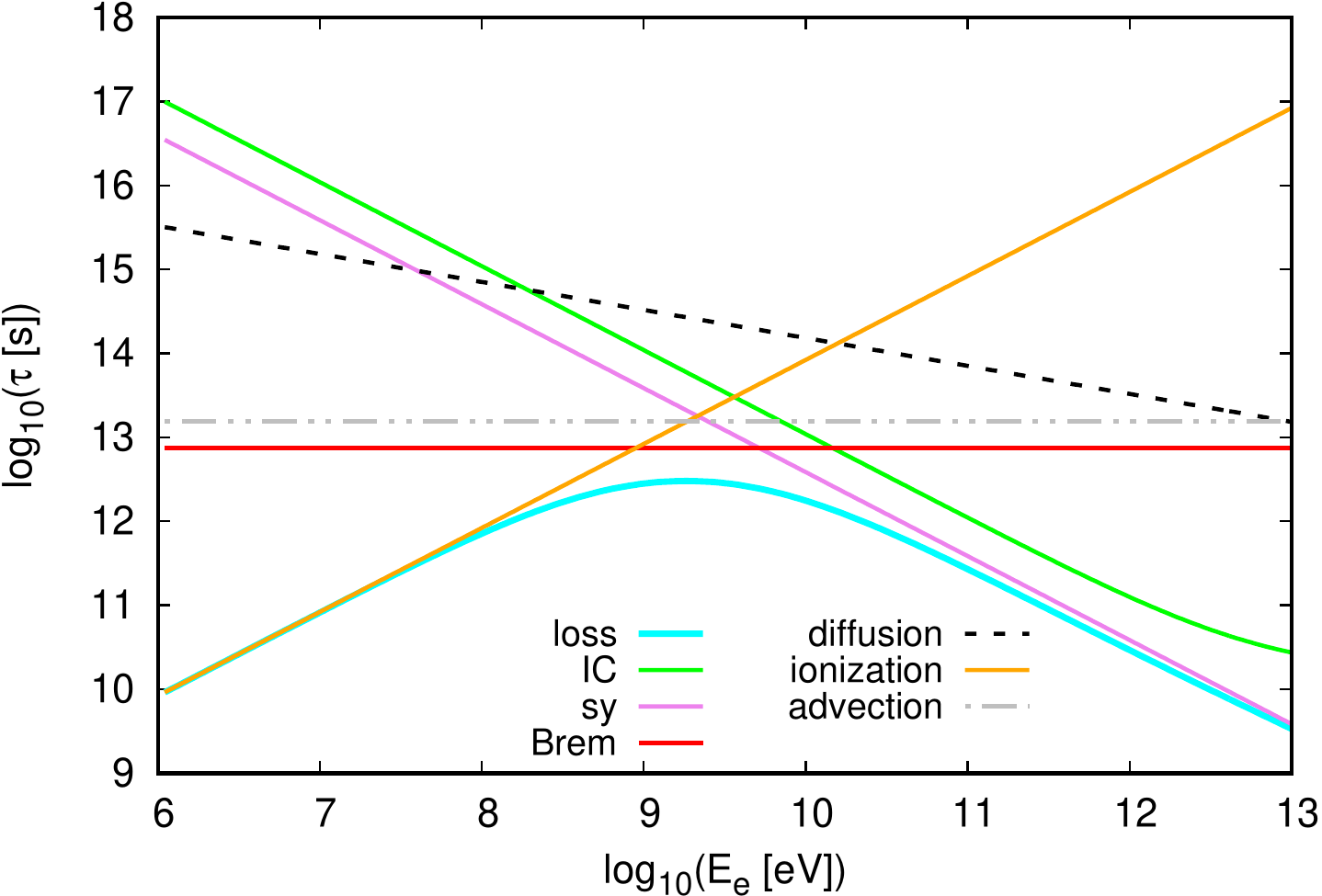}
    \includegraphics[angle=-90, width=0.49\textwidth]{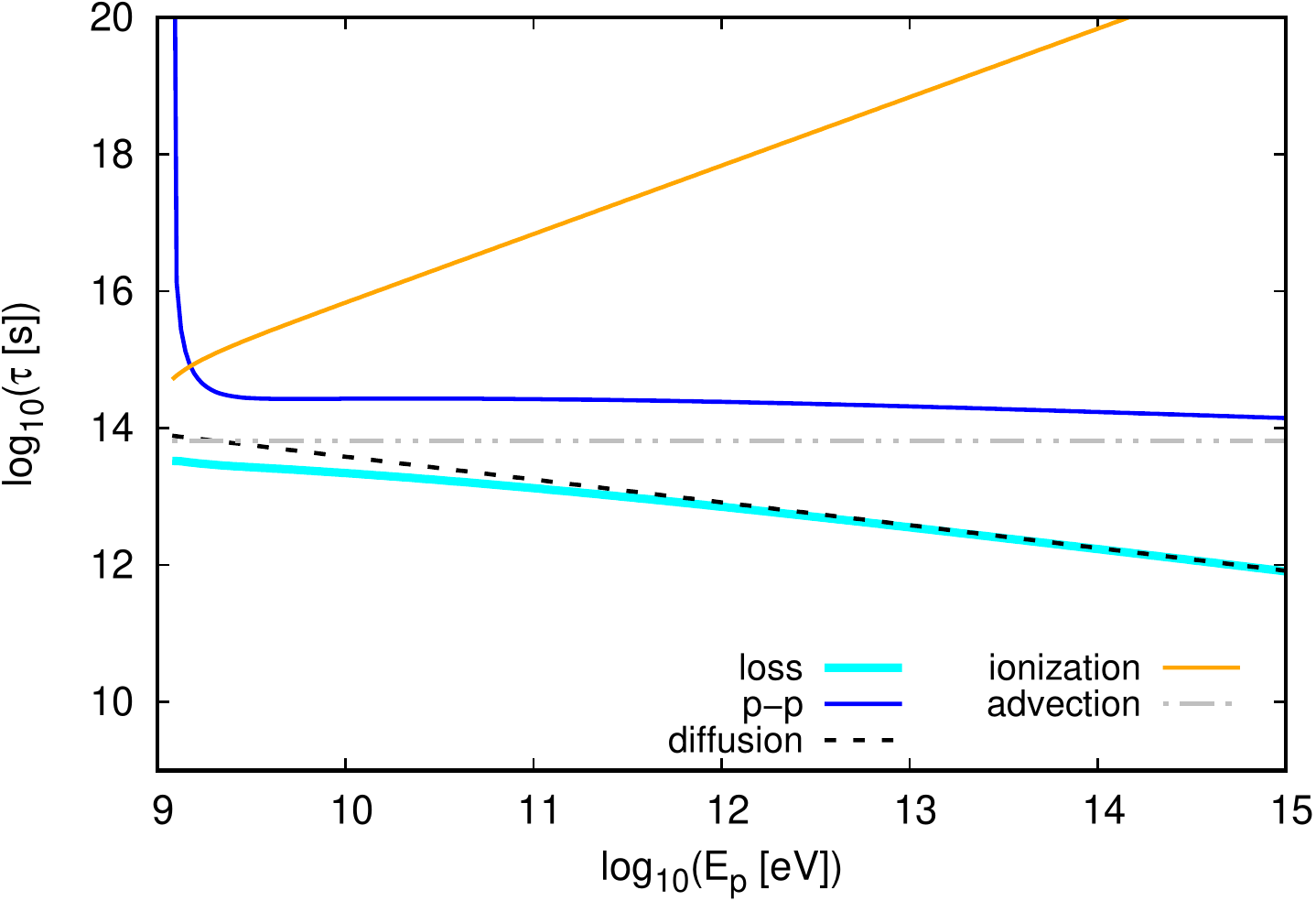}
    \includegraphics[angle=-90, width=0.49\textwidth]{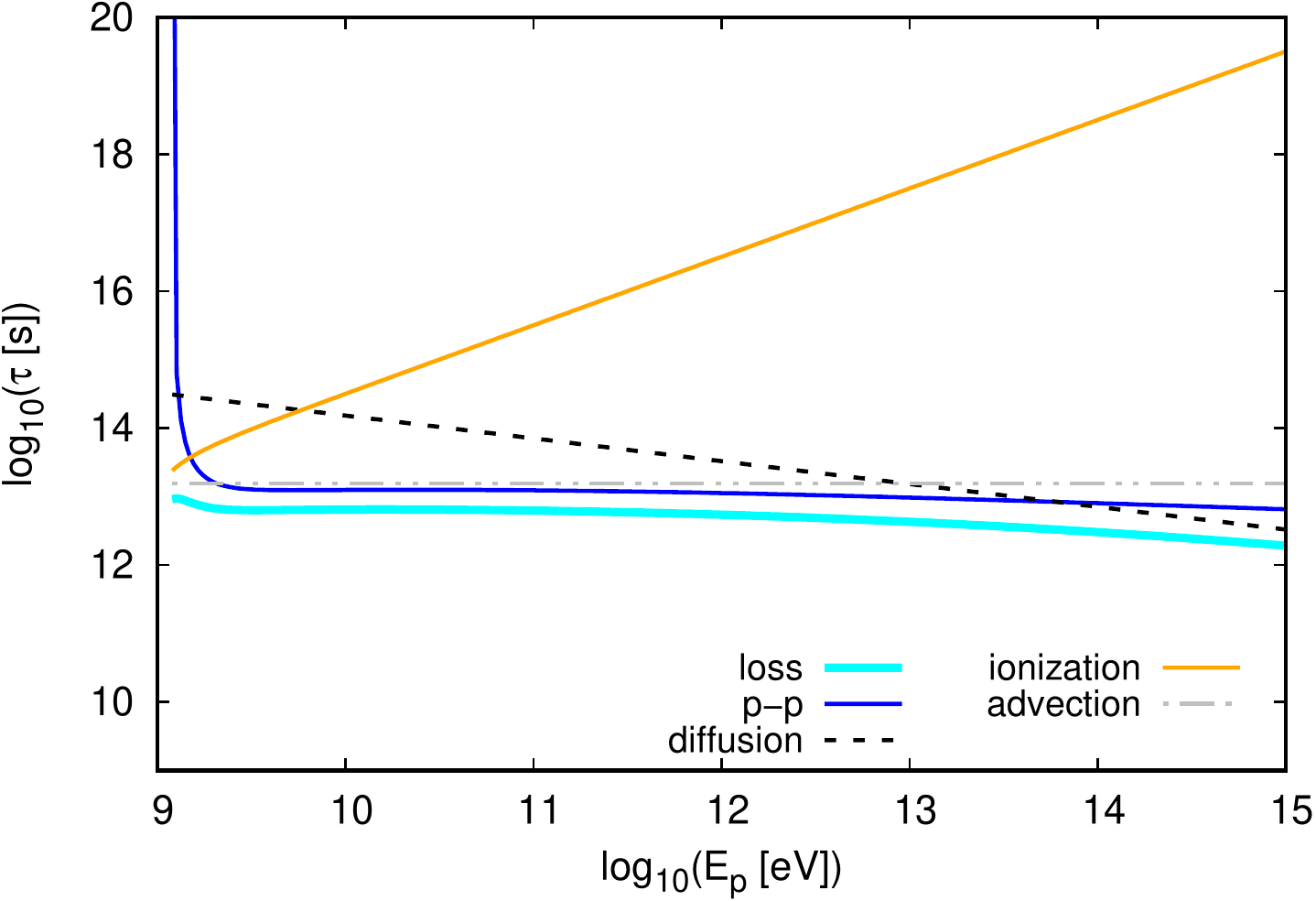}
    \caption{Cooling and escape times for primary electrons (\textit{upper panels}) and protons (\textit{bottom panels}) as a function of energy in scenario (0). The left panels show a typical case of a low SFR galaxy ($\dot{M}_*$ = 0.1~$M_{\odot}/\mathrm{yr}$). The right panels show a typical case of a high SFR system ($\dot{M}_*$ = 10~$M_{\odot}/\mathrm{yr}$). Solid lines represent different cooling processes: synchrotron (pink), ionisation (yellow), Bremsstrahlung (red), IC (green), and $p$-$p$ (blue). The black dashed lines are for diffusion, whereas the grey double-dot-dashed lines are for advection.}
    \label{fig: cooling}
\end{figure*}

The transport of electrons and protons in SFGs is strongly influenced by the average properties of the ISM, and in turn by their SFR. 
In this section we present the main transport properties of the benchmark scenario throughout the SFR range, analysing separately electrons and protons.
In order to highlight the evolution of the transport condition with the SFR, we show in the upper (lower) panels of Fig.~\ref{fig: cooling} the timescales of the relevant processes for electrons (protons) at two different values of the SFR: 
0.1~$\rm M_{\odot} \, yr^{-1}$ and 10~$\rm M_{\odot} \, yr^{-1}$.

\paragraph{Electrons.} Ionisation losses dominate the transport of electrons with $E_\mathrm{e} < 0.1$~GeV throughout the whole SFR range, while at higher electron energies the dominant loss mechanism depends on the SFR.

In the low SFR range, $\Dot{M}_* \lesssim 1 \, \rm M_{\odot} \, yr^{-1}$, electrons with energies between $0.5~\mathrm{GeV}$ and $ 10~\mathrm{GeV}$ escape efficiently from galaxies, whereas synchrotron confines electrons at higher energies.
As the SFR increases, Bremsstrahlung losses compete with the escape at $E_\mathrm{e} \sim 1$~GeV, while synchrotron keeps dominating the high energy domain together with IC scattering. 
In particular, the latter increases its relative importance for increasing SFR and eventually becomes the dominant cooling mechanism for SFR $\gtrsim 10^2 \, \rm M_{\odot} \, yr^{-1}$. 
Electron calorimetry is achieved for $\Dot{M}_* \gtrsim 5 \, \rm M_{\odot} \, yr^{-1}$, at which Bremsstrahlung dominates the cooling of GeV electrons; higher energy electrons are cooling-dominated for all SFR ranges. 

\paragraph{Protons.}
The transport of protons is regulated by the competition between escape (diffusion and advection) and energy losses ($p$-$p$ collisions). 
At low SFR, $\Dot{M}_* \lesssim 1 \, \rm M_{\odot} \, yr^{-1}$, most of the protons escape via diffusion for $E_\mathrm{p} \gtrsim  10 \, \rm GeV$ and via advection at lower energies.
At higher SFR, up to a few tens of $\rm M_{\odot} \, yr^{-1}$, $p$-$p$ losses become the dominant loss mechanism. However calorimetry is not completely achieved because the advection and $p$-$p$ timescales remain comparable. 
A transition to a diffusive-dominated transport also occurs in the multi-TeV range. 
Finally, we note that galaxies behave as good proton calorimeters for SFR $\gtrsim 30 \, \rm M_{\odot} \, yr^{-1}$.

\subsection{The $L_{\gamma}$--SFR relation}
 \label{subsec: result_gamma-sfr}
 
\begin{figure}
  \includegraphics[width=0.45 \textwidth]{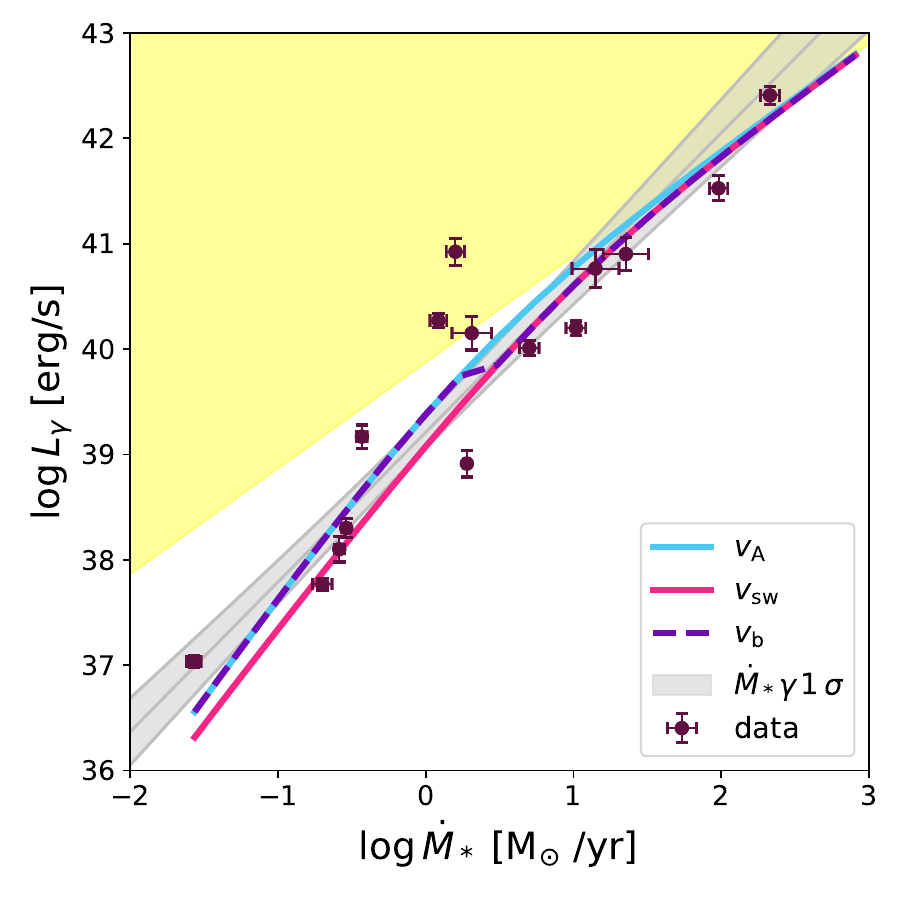}
  \caption{$L_{\gamma}$--SFR relation for scenario (0) using three different prescriptions for $v_\mathrm{adv}$: $v_\mathrm{a}$ (solid blue line), $v_\mathrm{sw}$ (solid pink line), and $v_\mathrm{b}$ from Eq.~\ref{eq:v_b} (violet dashed line). The grey line and shaded band indicate the best fit to the data (purple circles), and its $1 \sigma$ confidence region, respectively. The yellow region is forbidden by the calorimetric limit.}
  \label{fig: LgammaSFR}
\end{figure}

We define $L_{\gamma}$ as the integrated luminosity in the 0.1--100 GeV energy band, and we show in Fig.~\ref{fig: LgammaSFR} the $L_{\gamma}$--SFR relation obtained for the benchmark scenario. 
The main contribution to $L_{\gamma}$ comes from the $p$-$p$ interaction, while $\gamma$-rays of leptonic origin are subdominant (see Fig.~\ref{fig: SED}). 
In agreement with K20, the shape of the $L_{\gamma}$--SFR relation at low SFR is influenced by the escape of most of the injected protons,
whereas at high SFR sources approach the calorimetric limit, where the correlation becomes a linear scaling. 
We also show the results obtained assuming a unique nature of the advection speed throughout the whole SFR range, namely $v_{\rm adv}=v_\mathrm{A}$ or $v_{\rm adv}=v_\mathrm{sw}$.
Since $v_\mathrm{sw} > v_\mathrm{A}$ ($v_\mathrm{A} \gtrsim 50$~km\,s$^{-1}$ for the fiducial scenario), the CR escape is more efficient when $v_{\rm adv}=v_\mathrm{sw}$ and consequently the $\gamma$-ray luminosity is reduced. 
Our benchmark case is in good agreement with the best fit of the observed sample.

Variations of $\beta$, $B_0$ and $T_\mathrm{e}$ do not affect the emission of protons and, consequently, the $L_{\gamma}$--SFR correlation is not altered. 
On the contrary, the variation of the ionisation fraction $\rho$ (scenarios (5), (6)) modifies the CR escape through $v_\mathrm{A}$ at low SFRs, affecting the total source luminosity. 

As described in Sec.~\ref{subsec: free-free}, we explore values for the ionisation fraction down to $10^{-4}$.
While these values can characterise powerful starbursts such as ULIRGs, they are unlikely to be physical in standard SFGs with low SFR. Indeed, such low ionisation fraction leads to an Alfv\'en speed larger than the average velocity of starburst superwinds resulting in an advection dominated transport at odd with observations of CRs \citep{Blasi-review} and $\gamma$-rays \citep{ajello2020}.
The scenario $\rho=0.1$ does not show any relevant modifications with respect to the benchmark scenario since the dominant escape mechanism is diffusion.

As mentioned in Sec.~\ref{Sec: Model and Scale relations}, we improved the model of K20 introducing a SFR-dependent diffusion coefficient  and
a two-fold nature of the wind, while both diffusion coefficient and the wind velocity do not depend on the SFR in K20. Despite of these differences, the results presented here are compatible with the benchmark scenario discussed in K20.

The $L_{\gamma}$--SFR relation obtained here is in good agreement with the trend of data, and consistent with the results presented in K20. The model approaches the calorimetric limit (limit of the yellow shaded region) for $\dot{M}_* \geq 30 \, M_{\odot} \rm yr^{-1}$, above which galaxies show a calorimetric behaviour. NGC~3424 and 4945 luminosities are incompatible with this limit by 4.1 and 6.4 times their errors. As discussed in K20, the former probably harbours a low-luminosity AGN, whereas the SFR of the latter may have been underestimated due to its obscuration and inclination to the line of sight.

\subsection{The $L_{\mathrm{1.4 GHz}}$--SFR relation}
 \label{subsec: result_1.4-sfr}

\begin{figure*}
  \includegraphics[width=0.45 \textwidth]{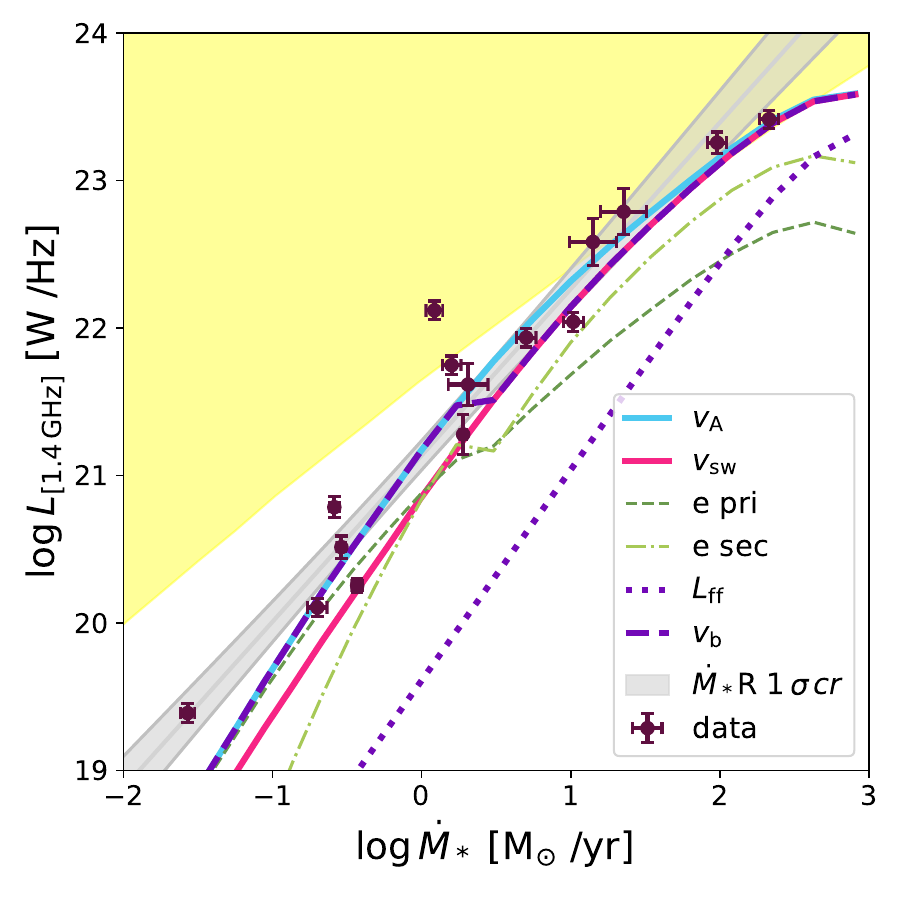}
  \includegraphics[width=0.45
  \textwidth]{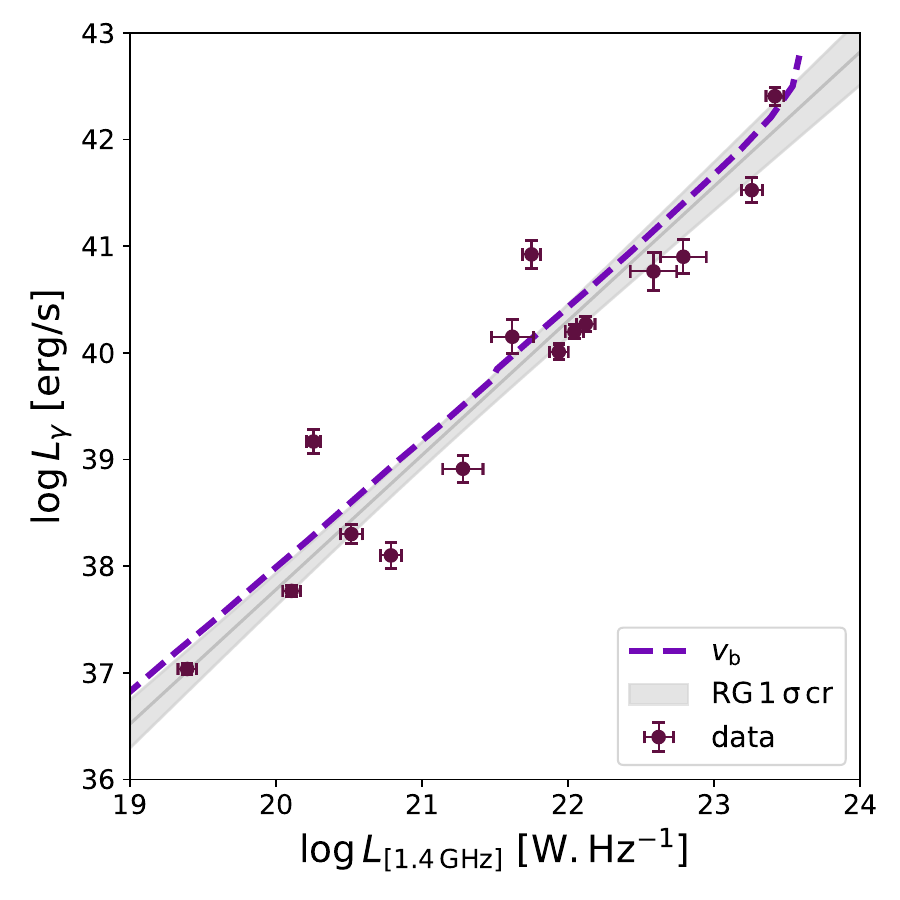}
  \caption{Results for scenario (0) using  $v_\mathrm{adv} = v_\mathrm{b}$ from Eq.~\ref{eq:v_b} (violet dashed line). \textit{Left panel}: The $L_{\mathrm{1.4 GHz}}$--SFR relation. The green dashed and dot-dashed lines represent the primary and secondary synchrotron electron contribution, respectively, and the dotted violet line the free-free contribution. The total emission at $L_{\mathrm{1.4 GHz}}$ is also shown for two different prescriptions for $v_\mathrm{adv}$, $v_\mathrm{A}$ (solid blue line) and $v_\mathrm{sw}$ (solid pink line). \textit{Right panel:} The $L_{\gamma}$--$L_{\mathrm{1.4 GHz}}$ relation. In both panels the grey shaded band indicates the $1 \sigma$ confidence region of each fit (solid grey line) to the data sample (purple points).}
  \label{fig_model_1_4}
\end{figure*}

The emission at 1.4~GHz is set by three processes: synchrotron radiation from both primary and secondary electrons, free-free (thermal) emission of the ionised gas, and free-free absorption produced by this same gas. Synchrotron radiation dominates the $L_{1.4 \mathrm{GHz}}$ throughout the entire SFR range, as shown in Fig.~\ref{fig_model_1_4} (left panel), while free-free emission and absorption become relevant only at high SFR. Eventually the thermal contribution becomes comparable to that of synchrotron at $\dot{M}_* > 100 \, \rm M_{\odot} \, yr^{-1}$ (see Fig~\ref{fig: SED}).

The emission of primary electrons dominates the $L_{1.4 \rm GHz}$--SFR correlation in the low SFR range ($\lesssim 1 \, \rm M_{\odot} \, yr^{-1}$), whereas synchrotron from secondaries gradually increases with the SFR and overcomes that of primaries at high SFR (see respectively green dot--dashed and dashed lines in left panel of Fig.~\ref{fig_model_1_4}). Such a behaviour is a direct consequence of the transport condition of the parent protons. In particular, the higher the SFR, the higher the gas density and so the rate of interaction of protons.

The secondary--to--primary ratio increases with the SFR as a consequence of the increasing level of calorimetry up to $\gtrsim 30 \, \rm M_{\odot} \, yr^{-1}$, where full calorimetry is achieved and the secondary--to--primary ratio becomes constant. This result is found to be in agreement with those of \citet{Peretti2019}.

In SFGs with SFR $\lesssim 10^2 \, \rm M_{\odot} \, yr^{-1}$, $L_{1.4 \rm GHz}$ is dominated by the synchrotron emission, hereafter $L_{\rm syn}^{1.4}$. In order to understand the shape of the correlation, we note that $L_{\rm syn}^{1.4}$ is dominated by the emission of $\sim$ GeV electrons, independently of the SFR. This electron-energy band is characterised by the competition of several cooling and escape mechanisms, so that the slope of the particle energy distribution can range from $\alpha-1$ when ionisation dominates to $\alpha+1$ when synchrotron dominates.
In addition, secondary electron emission depends on the SFR through both the density of the medium and the parent proton transport (see Sec.~\ref{SubsecSec: Seconday}). This leads to a complex dependence of $L_{\rm syn}^{1.4}$ on the SFR, which we parametrise as 
\begin{equation}
    \label{eq: Lsyn_dep}
    L_{\rm syn}^{1.4}(\dot{M}_*) \propto \dot{M}_*^{\eta(\dot{M}_*)} ,
\end{equation}

\noindent
where $\eta$ is the slope of the synchrotron luminosity in the logarithmic $L_{1.4 \rm GHz}$--SFR correlation. Notice that such component dominates the $L_{1.4 \rm GHz}$--SFR correlation up to $\sim 10^2 \,\rm M_{\odot} \, yr^{-1}$ for the benchmark scenario, where we constrain $\eta$ to the range 0.8--1.8.
In particular, the slope of the correlation is steeper in the low SFR range, where the relative contribution of secondaries increases with the SFR. 
At high SFR, also the transport of protons becomes calorimetric and the slope of the correlation softens (see Appendix~\ref{ap:model} for a detailed discussion). Notice that in calorimetric conditions $\eta < 1$. This results from the higher efficiency of ionisation and Bremsstrahlung in cooling GeV electrons.

The relevance of thermal processes increases for higher SFR as a consequence of the higher density of the ISM and our assumption of constant ionisation fraction $\rho$ throughout the whole SFR range. In particular, the thermal contribution to $L_{1.4\mathrm{GHz}}$ is at the level of $\sim 5\%$ for low SFR galaxies; this is consistent with values reported by \citet{Condon1992} for the thermal emission in the MW and M82. At high SFRs ($\gtrsim 100 \, \rm M_{\odot} \, yr^{-1}$) the thermal contribution is $\sim 50 \%$, and the free-free absorption also becomes significant. 
Nonetheless, assumptions in the ionisation fraction $\rho$ impact on the relative importance of the free-free processes, as we will discuss below.

Finally, we compare our benchmark model with the data presented in Table~\ref{tabledata}. We highlight that the coexistence of both primary and secondary synchrotron electron components is necessary to reproduce the slope of the $L_{1.4 \rm GHz}$--SFR correlation. 
On the right panel of Fig.~\ref{fig_model_1_4} we present observational data in the $L_{\gamma}$--$L_{1.4 \rm GHz}$ plane and compare them with the outcomes of our model; 
we obtain a good agreement throughout the whole SFR range.

\subsection{Parameter dependence of the $L_{\mathrm{1.4 GHz}}$--SFR relation}
 \label{subsec: param_1.4-sfr}

In this section, we discuss the impact of parameter variations on the $L_{1.4 \rm GHz}$--SFR correlation.

\begin{figure*}
  \centering
  \includegraphics[width=0.41 \textwidth]{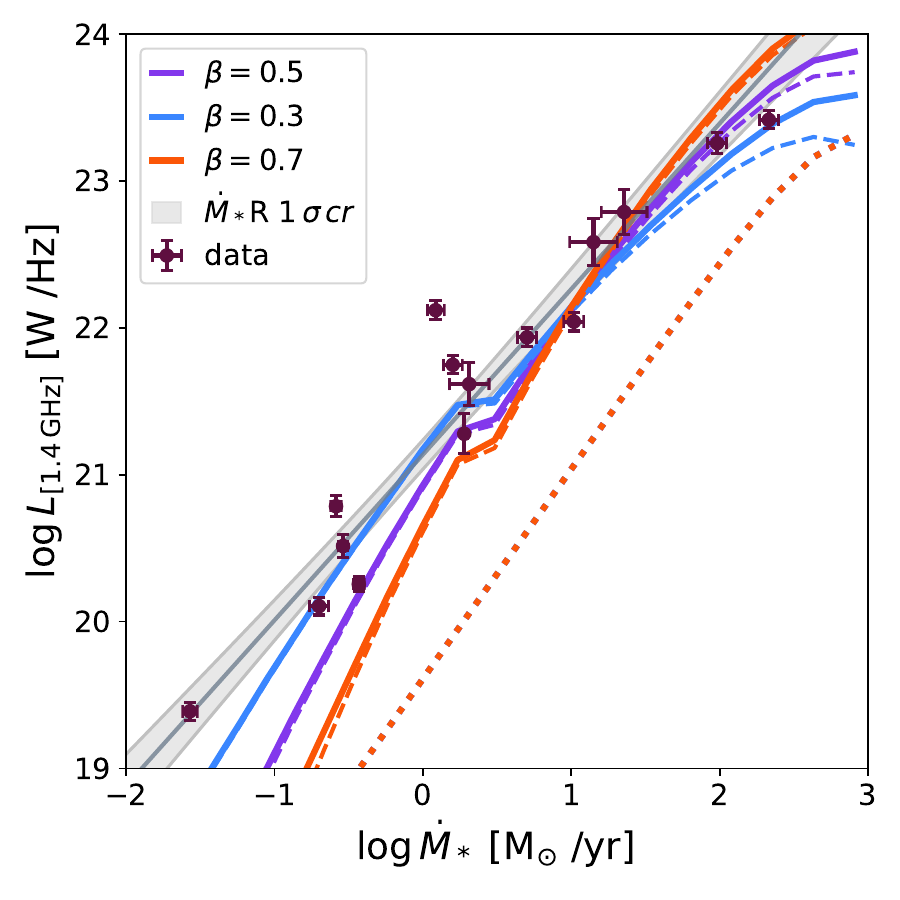}
  \includegraphics[width=0.41 \textwidth]{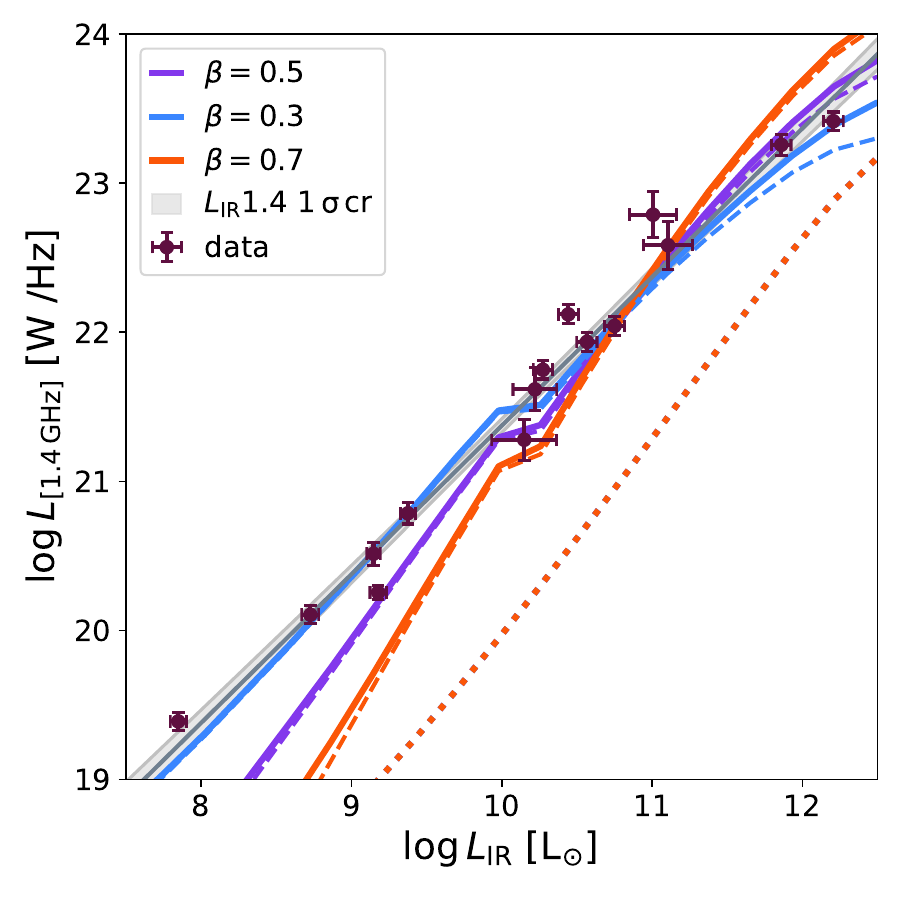}\\
  \includegraphics[width=0.41 \textwidth]{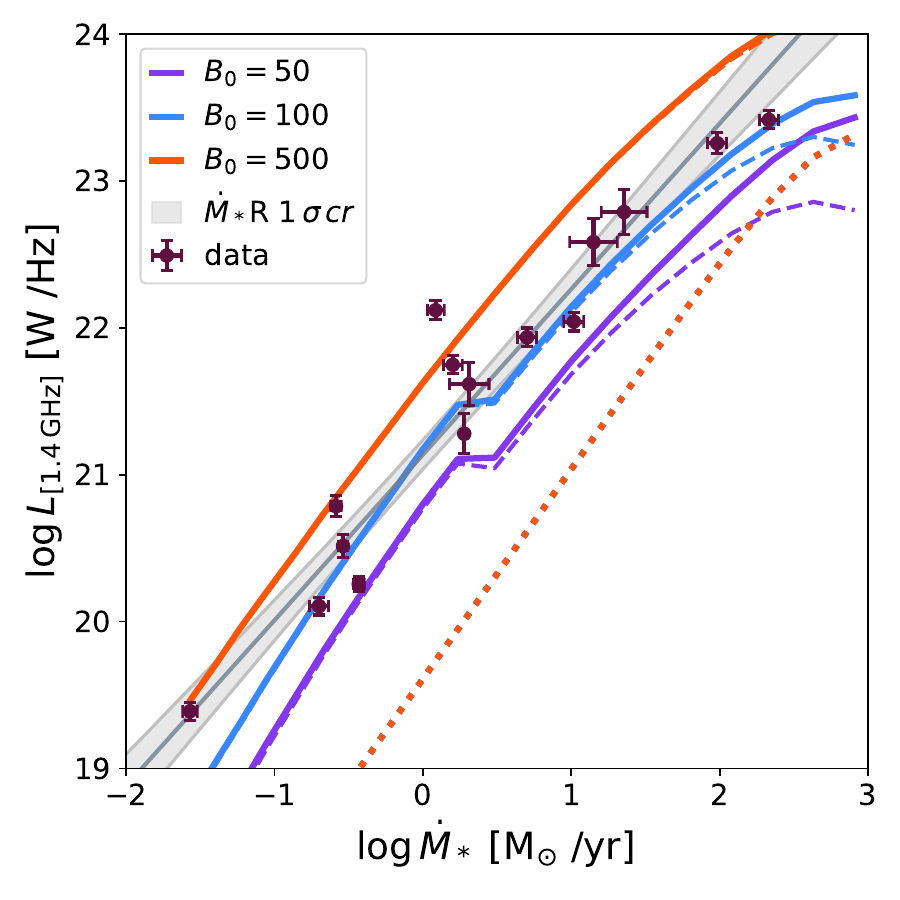}
  \includegraphics[width=0.41 \textwidth]{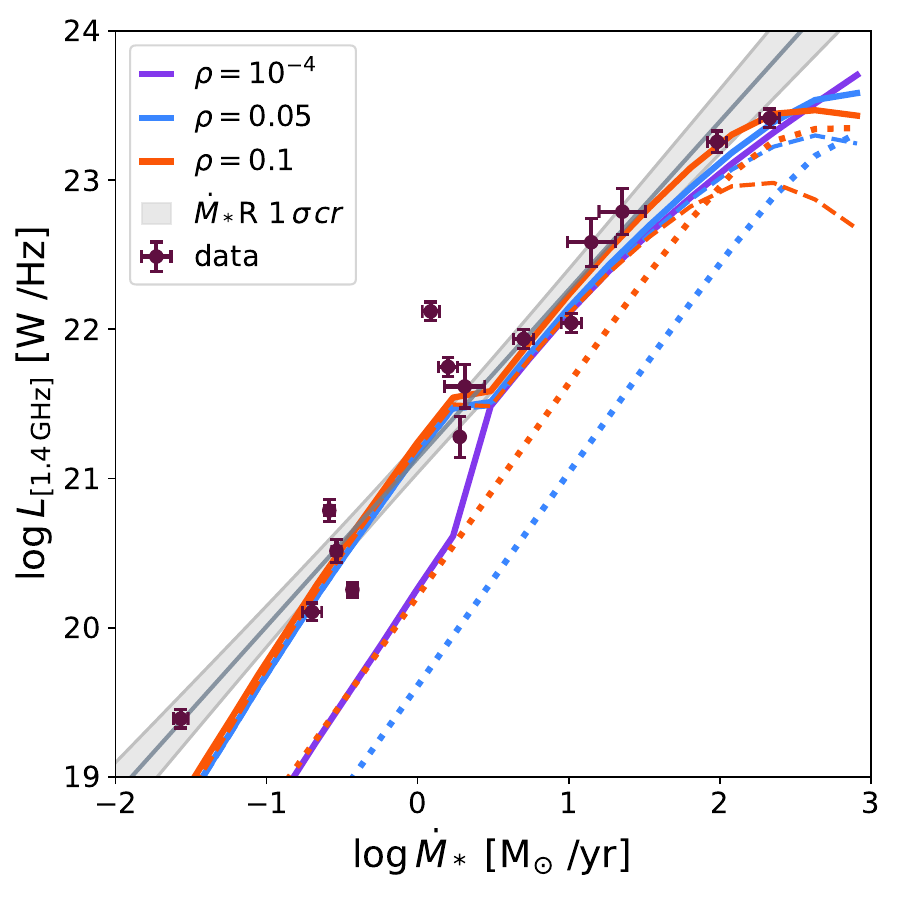}\\
  \includegraphics[width=0.41
  \textwidth]{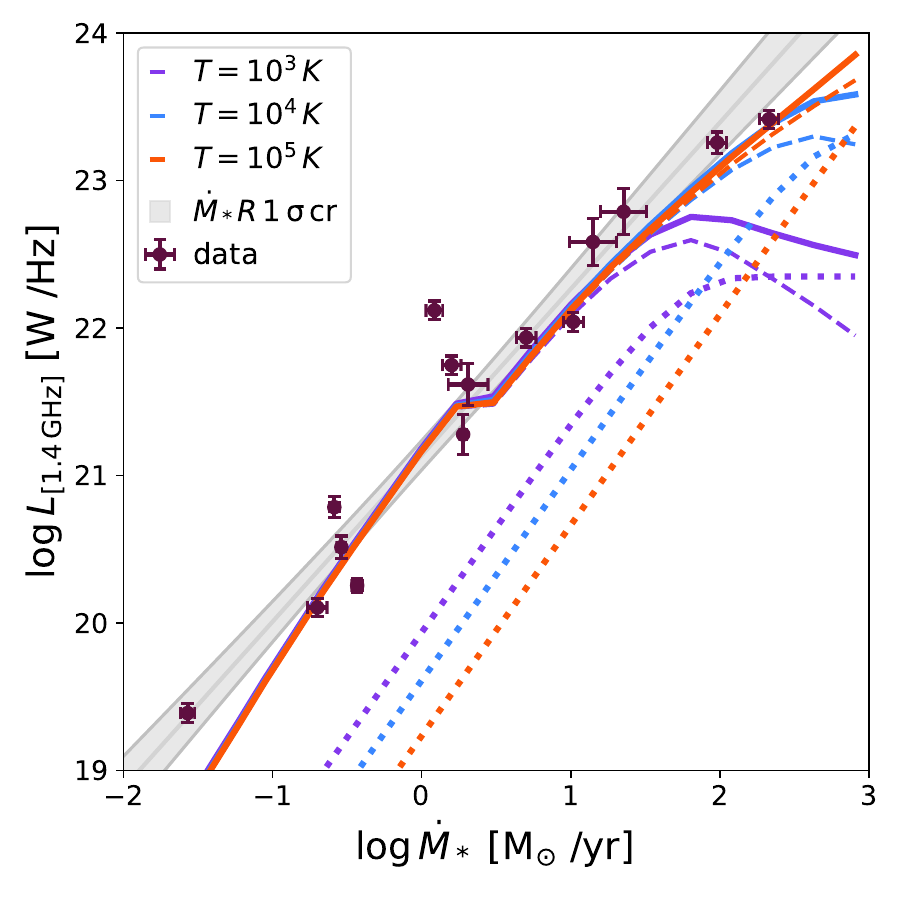}
  \includegraphics[width=0.41
  \textwidth]{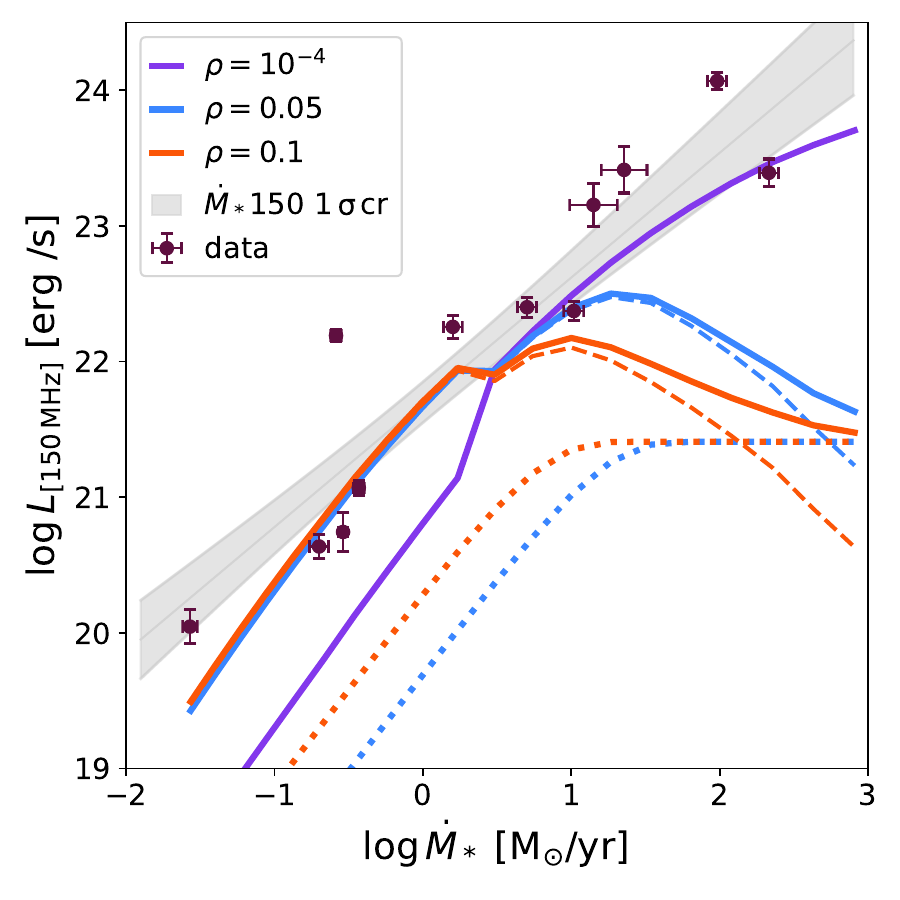}
\caption{Model relations for the different scenarios considered (Table~\ref{Table_modelparameters}).
The grey shaded band indicates the $1 \sigma$ confidence region of each fit (grey solid line) to the data sample (purple points). In all panels the dotted lines represent the thermal emission, the dashed lines the non-thermal emission, and the solid line the total emission. \textit{Upper panel}: for scenarios (0), (1) and (2) 
we show the $L_{\mathrm{1.4 GHz}}$--SFR relation (left) and the $L_{\mathrm{1.4 GHz}}$--$L_\mathrm{IR}$ relation (right). \textit{Middle panel}: We show the $L_{\mathrm{1.4 GHz}}$--SFR relation for scenarios 0, 3 and 4 (left) and (0), (5) and (6) (right). \textit{Bottom panel:} for scenarios (0), (7) and (8) we show the $L_{\rm 1.4 GHz}$--SFR relation (left) and for scenarios (0), (5) and (6) the $L_{\mathrm{150 MHz}}$--SFR relation (right). }
  \label{fig_var_par}
\end{figure*}

\paragraph{Magnetic field index.} As the thermal emission is unaltered by changes in $\beta$, hereafter we focus only on the synchrotron component.  
Top panels in Fig.~\ref{fig_var_par} illustrate how the variation of $\beta$ (scenarios (1) and (2)) affects the slope of the $L_{\mathrm{1.4 GHz}}$--SFR and $L_{\mathrm{1.4 GHz}}$--$L_{\rm IR}$ correlations. 

In particular, the larger the value of $\beta$, the steeper the slope $\eta$ of the $L_{1.4 \rm GHz}$--SFR correlation. 
This is a natural consequence of the dependence of the synchrotron luminosity on the magnetic field which, for larger $\beta$, presents a larger variation throughout the SFR range (see Appendix~\ref{ap:model}). 

At the same time, lower values of $B$ resulting from $\beta>0.3$ at low SFR reduce the escape time, which in turn reduces the luminosity and makes the correlation steeper.
The benchmark scenario with $\beta = 0.3$ provides the best fit to the general trend of the observed data.
However, the assumption $\beta = 0.5$, in agreement with \citet{Lacki2010}, looks somehow more appropriate while fitting data at high SFR.
The scenario assuming $\beta =0.7$
is disfavoured when compared with data in the entire SFR range whereas, as in the case of $\beta=0.5$, it can be adopted to describe the high SFR range.

\paragraph{Magnetic field normalisation.} A variation of $B_0$ (scenarios (3), (4)) 
results in a different normalisation of the correlation curve of about a factor $\sim 1.5 - 4$. We show these results in the middle left panel of Fig.~\ref{fig_var_par}. This behaviour is in agreement with the dependence of the synchrotron luminosity of a non-thermal electron population with the magnetic field \citep[see e.g.][]{Ghisellini2013}.
However, the synchrotron radiation at 1.4 GHz is mostly produced by electrons in the $\sim$ GeV range. At these energies several physical processes compete in shaping the electron energy distribution. 
This leads to an energy distribution different from a simple power law. Thus, in the context of SFGs, the dependence of $L_\mathrm{1.4GHz}$ on the magnetic field strength is not trivial, and a change in the normalisation constant $B_0$ introduces an overall shift which slightly depends on the SFR. 
Despite this complexity, we show that the parameter $B_0$ does not affect the global slope of the correlation but only influences its normalisation (see Appendix~\ref{ap:model} for a more detailed and quantitative discussion). 
The value $B_0 = 100$ is the one that best fits the data. The resulting correlations for $B_0 = 50 $ and 500, limits the data above and below respectively, excluding only \object{NGC 4945}.

\paragraph{Ionisation fraction.} The impact of different assumptions for the ionisation fraction $ \rho$ on the $L_{1.4 \rm GHz}$--SFR correlation is threefold since it affects: 1) the free-free emission, 2) the free-free absorption and 3) the advection timescale in low SFGs. 
The middle right panel of Fig.~\ref{fig_var_par} illustrates the results obtained in the benchmark case (blue lines) compared with $\rho=10^{-4}$ (violet lines) and $\rho=10^{-1}$ (orange lines).  

Although a value of $\rho$=0.1 produce results consistent with the observed trend, at high SFR the thermal emission becomes significant and possibly dominates. This is the opposite of what is expected for galaxies like M82 where the estimated thermal fraction is even lower \citep[$\sim 2 \%$,][]{Basu2012} than for normal galaxies.
The scenario where $\rho = 0.1$ is therefore unlikely to be realised in nature and at odds with multi-wavelength modelling of ULIRGs \citep[see e.g.][]{Yoast-Hull-ARP}.
On the other hand, lower values of $\rho$ reduce the absorption and the relative importance of thermal emission at high SFR.
This yields a thermal contribution that is too low with respect to what is generally accepted to hold for normal galaxies \citep{Condon1992}. Finally,  $\rho = 10^{-4}$ underestimates the data at low SFR and the model behaviour is similar to that of $\gamma$-rays.

We conclude that the scenario where $\rho = 0.05$ is the one that better accommodates the observed data and remains consistent with synchrotron dominating the emission at 1.4~GHz.
Although it is tempting to speculate on a possible SFR--dependence of $\rho$, this would introduce further complexity in the model, and we therefore leave it for upcoming investigations.

\paragraph{Electron temperature.} $T_\mathrm{e} \sim 10^4$~K is the standard temperature assumed for the warm ionised ISM. 
In starburst regions, this gas coexists with a hot and pressurised component resulting from the collective effect of supernovae and stellar winds. This implies that $T_\mathrm{e} > 10^4$~K might be consistent with the average temperature in these environments \citep{Westmoquette2009}. 
The variation of $T_\mathrm{e}$ (scenarios (7), (8)) affects the thermal emission and absorption. 
Such behaviour is shown in Fig.~\ref{fig_var_par} (bottom left panel) from which it can be concluded that values between $T_\mathrm{e} = 10^4$ and $10^5$~K reproduce well the observations, whereas lower values of $T_\mathrm{e}$ lead to an excessive absorption at high SFR compared to the data trend.

These results can be understood by looking at the physical dependence of the opacity on the temperature. In particular,
at a given photon frequency, $\tau_\mathrm{ff} \propto T_{\mathrm{e}}^{-3/2} \dot{M}_*^{1.7}$, and so the optically thick regime is reached at higher SFRs for higher values of $T_\mathrm{e}$. 
Thus, the absorption of synchrotron radiation decreases with $T_\mathrm{e}$. 

Finally, we observe that the different assumptions about the temperature do not strongly impact the relative contribution of thermal emission to the $L_{1.4 \rm GHz}$ across the whole SFR range.
 
\subsection{The $L_{\mathrm{150 MHz}}$--SFR relation}
 \label{subsec: result_150-sfr}

\begin{figure}
  \includegraphics[width=0.45 \textwidth]{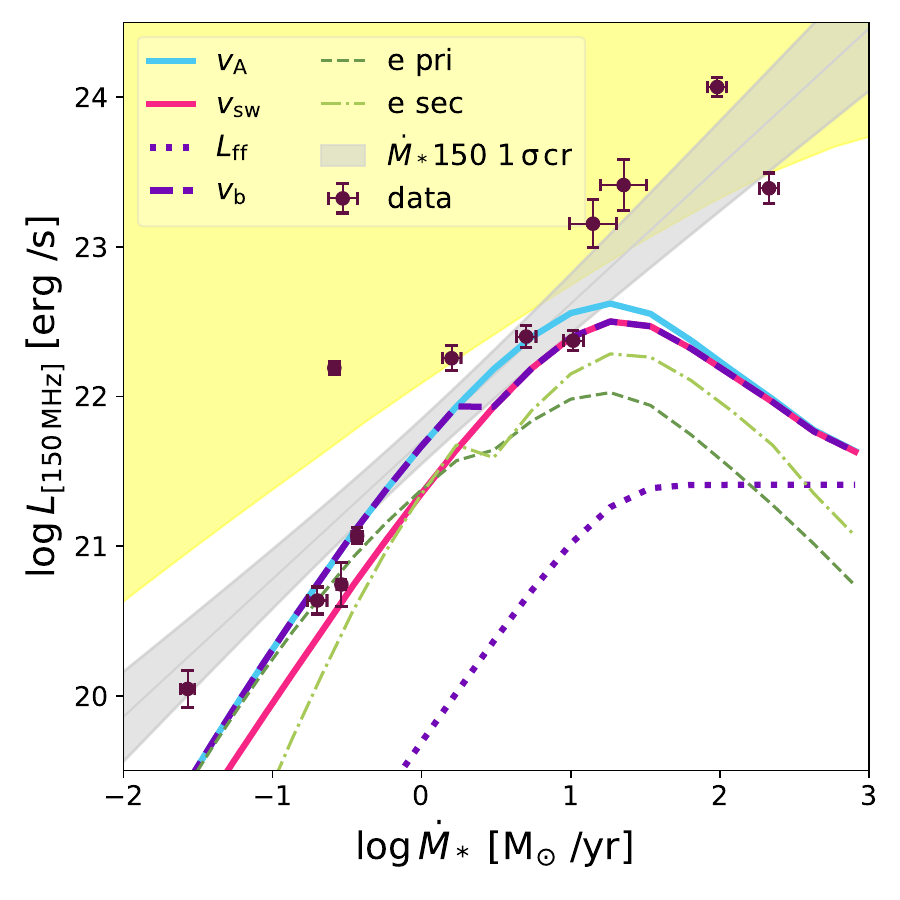}
  \caption{The $L_{\mathrm{150 MHz}}$--SFR relation obtained for scenario (0) using three different prescriptions for $v_\mathrm{adv}$: $v_\mathrm{A}$ (solid cyan line), $v_\mathrm{sw}$ (solid pink line), and $v_\mathrm{b}$ from Eq.~\ref{eq:v_b} (violet dashed line). The green dashed and green dot-dashed lines represent the primary and secondary electron contributions, respectively, and the dotted violet line represents the free-free contribution. The grey shaded band indicates the $1 \sigma$ confidence region of the fit solid grey line to the data sample (purple points).}
  \label{fig150MHz}
\end{figure}

The physical processes governing the emission at 150~MHz are the same as at 1.4~GHz (Sec.~\ref{subsec: result_1.4-sfr}).
However, at this frequency free-free absorption plays a crucial role. 
In fact, the turnover frequency where $\tau_\mathrm{ff} > 1$ increases up to 150~MHz for $\Dot{M}_* \sim 10~M_{\odot}/ \mathrm{yr}$. 
Therefore, the 150-MHz synchrotron emission is completely absorbed at high SFRs in our benchmark scenario. 
This is reflected in the $L_{150 \mathrm{MHz}}$--SFR correlation shown in Fig~\ref{fig150MHz}. 
This scenario departs considerably from the observed trend of data at high SFRs and cannot reproduce the emission of the ULIRGs. 
Such a strong absorption is the result of the ionisation fraction assumed, $\rho \sim 5 \%$.   
In addition, the calorimetric limit determined for the benchmark scenario also shows that it can hardly match the observed fluxes for SFR $\gtrsim 10^2 \, \rm M_{\odot} \, yr^{-1}$.

Although the effects of the variation of parameters in the $L_{150 \mathrm{MHz}}$--SFR are similar to the $L_{1.4 \rm GHz}$--SFR correlation, some details are required to figure out if it is possible to match the trend of data with some model prediction under different parametric assumptions.
 
The assumption $\beta = 0.5$, as in the case at 1.4~GHz, increases the slope of both the correlation and the calorimetric limit.
Moreover, ionisation losses are dynamically important for electrons emitting synchrotron at 150~MHz. 
In addition, different assumptions for $\rho$ and $T_\mathrm{e}$ can also mitigate or enhance the absorption as evidenced in the right bottom panel of Fig~\ref{fig_var_par}. 
As described above for 1.4~GHz, it is hard to find a unique parameter configuration that can accommodate the entire SFR range without being at odds with observations. 
In particular, we observe that the introduction of a dependence on the SFR for $\rho$ and/or $T_\mathrm{e}$ would provide a better fit at high SFR. Such a lack of emission at low radio frequencies was also reported by \citet{Schober2017} who found a similar result studying the correlation at 200~MHz.

%
\section{Conclusions} 

\label{Sec: conclusions}
%

We investigated how the transport of CRs, produced by sources related to the star formation, shapes the observed $L_{\gamma}$--SFR, $L_{1.4 \mathrm{GHz}}$--SFR and $L_{150 \mathrm{MHz}}$--SFR correlations for SFGs detected in $\gamma$-rays. We presented a flexible emission model for SFGs over a broad range of SFRs and compared the theoretical estimates with the most updated observational data available. The main results from our work are summarised as follows:

\begin{itemize}

    \item We provide homogeneous radio continuum fluxes and their errors computed directly from images at  1.4  GHz of  Stokes-I available in the  NRAO/VLASky continuum Survey catalogue for NGC  253,  NGC 1068, NGC 2146, NGC 2403, M82, Arp 220, Arp 299, NGC 3424. We also compute for the first time the fluxes and the errors for NGC 4945 and Circinus from the new 1.4 GHz  Stokes-I continuum image provided by The ASKAP Continuum Survey.
    \item We provide the first self-consistent data set at 150~MHz and 1.4~GHz for all galaxies detected in $\gamma$-rays. We present updated correlations of $L_{\gamma}$--SFR, $L_{1.4\, \rm GHz}$--SFR and $L_{150\, \rm MHz}$--SFR for these galaxies. 
    \item The trend followed by the $L_{1.4\, \rm GHz}$--SFR correlation observed in the sub-sample of SFGs detected at $\gamma$-rays agrees very well with that obtained from larger surveys of SFGs \citep{Yun2001, Bell2003}. 
    \item We revisited the $L_{\gamma}$--$L_\mathrm{1.4 GHz}$ correlation and found a steeper slope $m =1.26 \pm 0.10$ (with $L_\gamma \propto L_\mathrm{1.4 GHz}^m$) with respect to the one reported by \citet{Ackermann2012}. 
    \item We obtained a slope of $m =  0.98 \pm 0.13$ for the $L_\mathrm{150 MHz}$--SFR correlation (with $L_\mathrm{150 MHz} \propto \dot{M}_*^m$), which is marginally shallower than results reported by other authors (ranging from 1 to 1.3). 
    \item We developed a self-consistent model for CR transport and multi-wavelength emission suitable to interpret the luminosity--SFR correlation at different energy ranges over a very broad range of SFRs. 
    This model can reproduce reasonably well the observed $L_{\gamma}$--SFR, $L_{1.4 \mathrm{GHz}}$--SFR and $L_{1.4 \mathrm{GHz}}$--$L_{\gamma}$ correlations. 

    \item A constant ionisation fraction $\rho$ fails to reproduce the $L_{\rm 150\, MHz}$--SFR correlation throughout the whole SFR range. In particular, an ionisation fraction of $\sim 5\%$ at low SFRs is compatible with observations, whereas at high SFRs a value $<1\%$ is favoured to achieve the emission observed in ULIRGs. Indeed, small values of $\rho$ would be in agreement with the results of \citet{Valdes2005} who suggest that a significant fraction of the ionising photons may be absorbed by dust. Moreover, \citet{Krumholz2020} propose that typical ionisation fractions in starbursts should be $\sim 10^{-4}$ since the hot gas would be advected in the wind.

    \item A  scaling law $B = B_0 (\dot{M}_*/10\,\rm M_{\odot} \rm yr^{-1})^\beta$ is compatible with the observed data for $\beta = 0.3$, which suggests that the magnetic field strength is regulated by stationary MHD turbulence. Higher beta values are also compatible with data limited to the range of high SFR. The normalisation value that best describes the magnetic field at $10\,\rm M_{\odot} \rm yr^{-1}$ is of the order of $100 ~\mu$G.
    \item The coexistence of synchrotron components from both primary and secondary electrons is mandatory to reproduce the slope of the $L_{1.4 \mathrm{GHz}}$--SFR correlation. 
    Such a slope reflects the transport conditions of primary protons and could provide indications of a transition to a calorimetric regime.
    \item High temperatures of $10^4$--$10^5$~K for the ionised gas are more suitable to explain the emission of the ULIRGs. 
    \item Galaxies behave as good proton calorimeters for SFR $\gtrsim 30 \, \rm M_{\odot} \, yr^{-1}$. 
    Such a calorimetry can be assessed via observations in $\gamma$-rays. On the contrary, relativistic electrons, despite reaching also the calorimetric limit for SFR $\gtrsim 5 \, \rm M_{\odot} \, yr^{-1}$, are hard to assess given that their emission is not completely dominant for all SFRs in either the radio or the $\gamma$-ray band. 
    \item Different wind prescriptions depending on the SFR of the galaxy yield a good description of the $L_\gamma$--SFR correlation. In particular, an alfvenic wind is proposed to dominate in low-SFR galaxies, whereas a thermally-driven wind would better describe the outflow at high SFRs. 
\end{itemize}
The SMC deserves a more thorough discussion. It is the galaxy at the lower end of the SFR range, separated by far from the next one, and therefore may play an important role in model fitting. Its non-thermal luminosity is almost always underpredicted by our models. It might be argued that this might be the result of additional contributions from either unresolved point-like sources or from the halo. However, this seems very improbable, due to the short distance and extended nature of this galaxy. The discrepancy can be mitigated taking into account the uncertainty in the current SFR estimates for the SMC. We have used the K20 value ($0.027\,\mathrm{M}_\odot \, \mathrm{yr}^{-1}$) obtained from an H$\alpha$/IR proxy, sensitive to the SFR in a time interval of $\sim 10\,\mathrm{Myr}$, whereas it has been found that the SMC has a complex SFR history with recent bursts of up to ten times this value \citep{Harris2004}. An increase of the SFR of the SMC of 0.5--1~dex results almost always in a better agreement between observations and our $L_{1.4 \mathrm{GHz}}$-SFR and $L_{150 \mathrm{MHz}}$-SFR modelled correlations at low SFRs.

Enlarging the current sample of $\gamma$-ray emitting SFGs, and developing reliable techniques capable of distinguishing the emission associated with AGNi hosted by some galaxies of the sample is necessary to obtain a more accurate $L_{\gamma}$-SFR correlation.
A larger sample of galaxies detected in $\gamma$-rays would also yield a more accurate $L_{\gamma}$--$L_\mathrm{1.4 \rm GHz}$ correlation, which is important for two reasons. The first one is that it could provide an independent method for the classification of $\gamma$-ray galaxies, in addition to the standard one based on the spectral index determination. The second one is that it could be used to estimate the contribution of SFGs to the $\gamma$-ray background, by extrapolating the radio luminosity function to high redshift.

A careful study of the $L_{150 \rm MHz}$--SFR correlation is important to assess the extent to which $L_{150 \rm MHz}$ traces star formation. Furthermore, a more refined $L_{150 \rm MHz}$--SFR correlation
would allow to discern if the SFGs detected in $\gamma$-rays follow a distinguishable trend with respect to more general samples. Due to the difficulties in reproducing the emission at these low frequencies,  in upcoming studies we will address the effect produced on the modelled correlation by an ionisation fraction diminishing with the SFR, or a highly clumped gas distribution resulting in a porous medium. An alternative scenario can be a non-negligible contribution from particles radiating from a region larger than our galactic box, for instance electrons diffusing or advected in the halo \citep{Heesen2009}, or the presence of an additional source/acceleration region (such as a co-existing AGN) not directly related to the SFR \citep{Gurkan2018}.

The next generation of telescopes with better angular resolution and sensitivity will allow unravelling the morphology of the diffuse non-thermal emitting region in SFGs, possibly resolving the emission coming from the disc from that originated in the halo or wind \citep[see e.g.][]{Romero2018,Ana_vientos,Tano_vientos}. 
The panchromatic view of the SFR--luminosity correlations in SFGs is a key to assess the physical processes that govern the non-thermal emission of these sources. 
In particular, we plan to extend the study of luminosity correlations to the VHE and hard X-ray energy bands. The former can provide a unique identification tool for possible new detections by the upcoming Cherenkov Telescope Array \citep[CTA;][]{CTA2019}. 
The latter can provide a new powerful tool for investigating the CR transport properties.

\begin{acknowledgements}
      P.K. and P.B. acknowledge B. Koribalski for help in providing the ASKAP data. The research activity of E.P. was supported by Villum Fonden (project n. 18994) and by the European Union’s Horizon 2020 research and innovation program under the Marie Sklodowska-Curie grant agreement No. 847523 ‘INTERACTIONS’. P.B. acknowledges support from ANPCyT PICT 0773--2017. L.J.P, P.K. and P.B. acknowledge also support from CONICET grant PIP 2014-0265.
\end{acknowledgements}

%
   \bibliographystyle{aa} 
   \bibliography{ref} 
%
\begin{appendix}
\section{Radio images}
\label{ap:images}
\begin{figure*}
\label{Fig: images}
  \centering
    \includegraphics[angle=-90, width=0.38
  \textwidth]{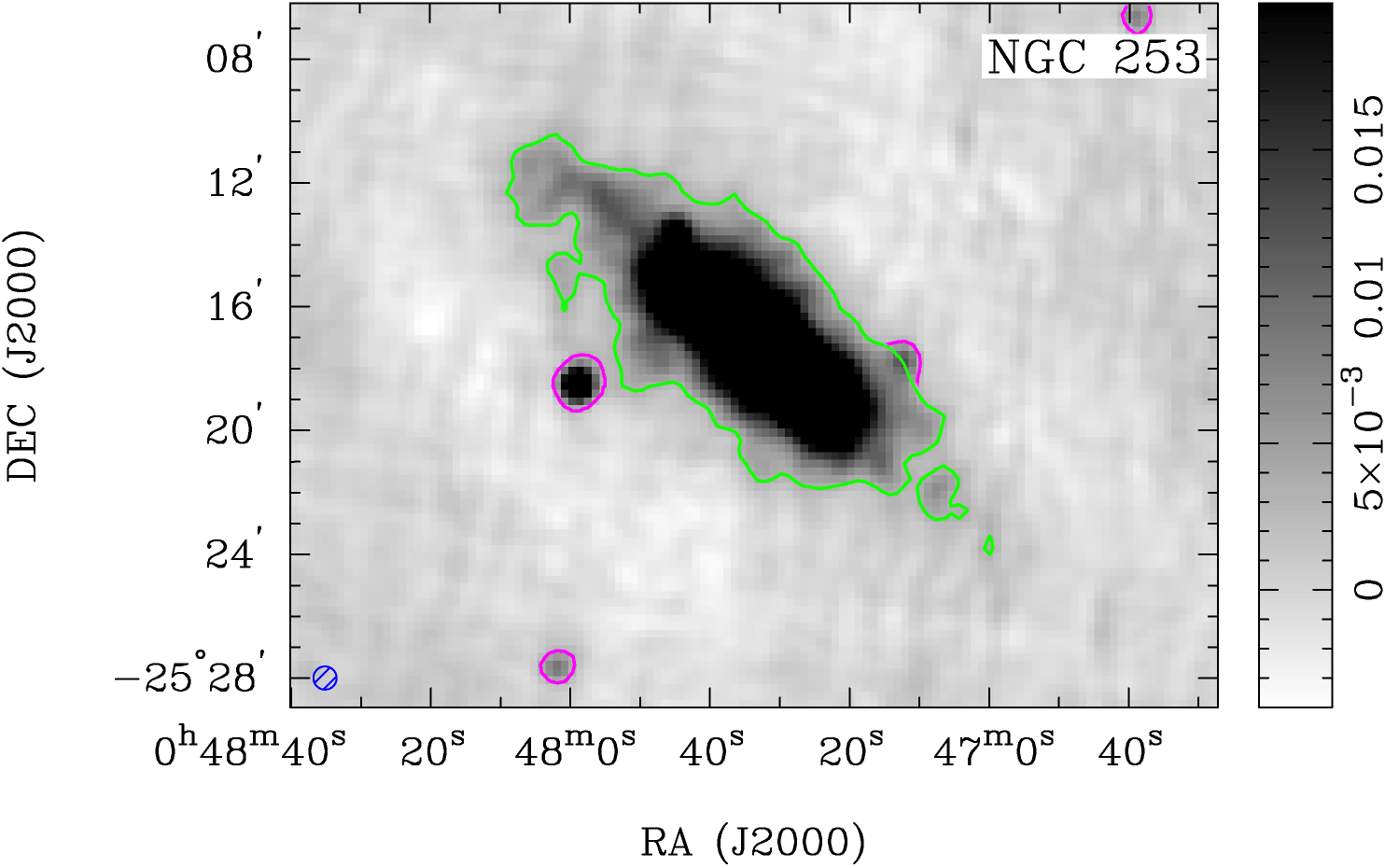}
    \includegraphics[angle=-90, width=0.38
  \textwidth]{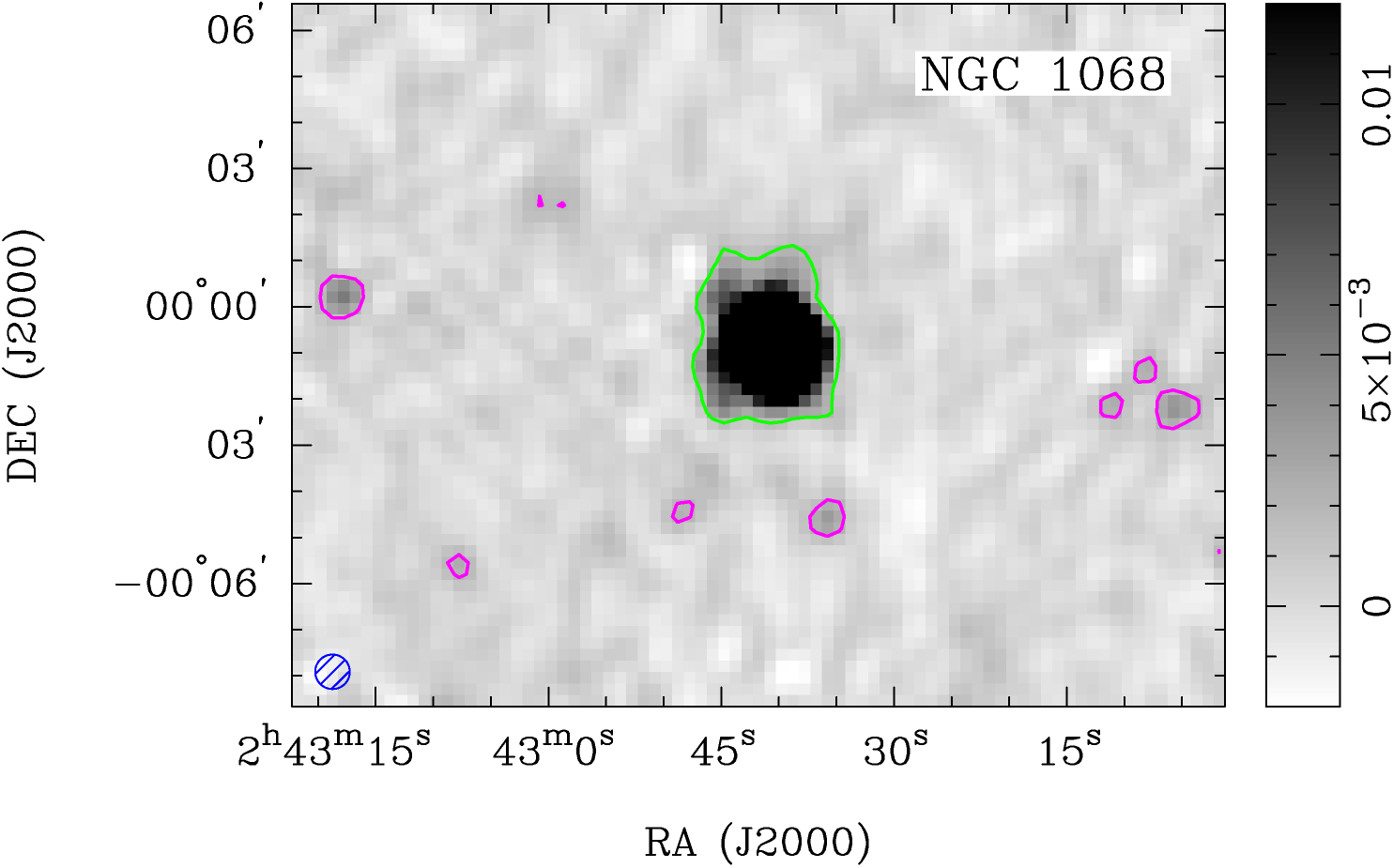} \\
  
      \includegraphics[angle=-90, width=0.38
  \textwidth]{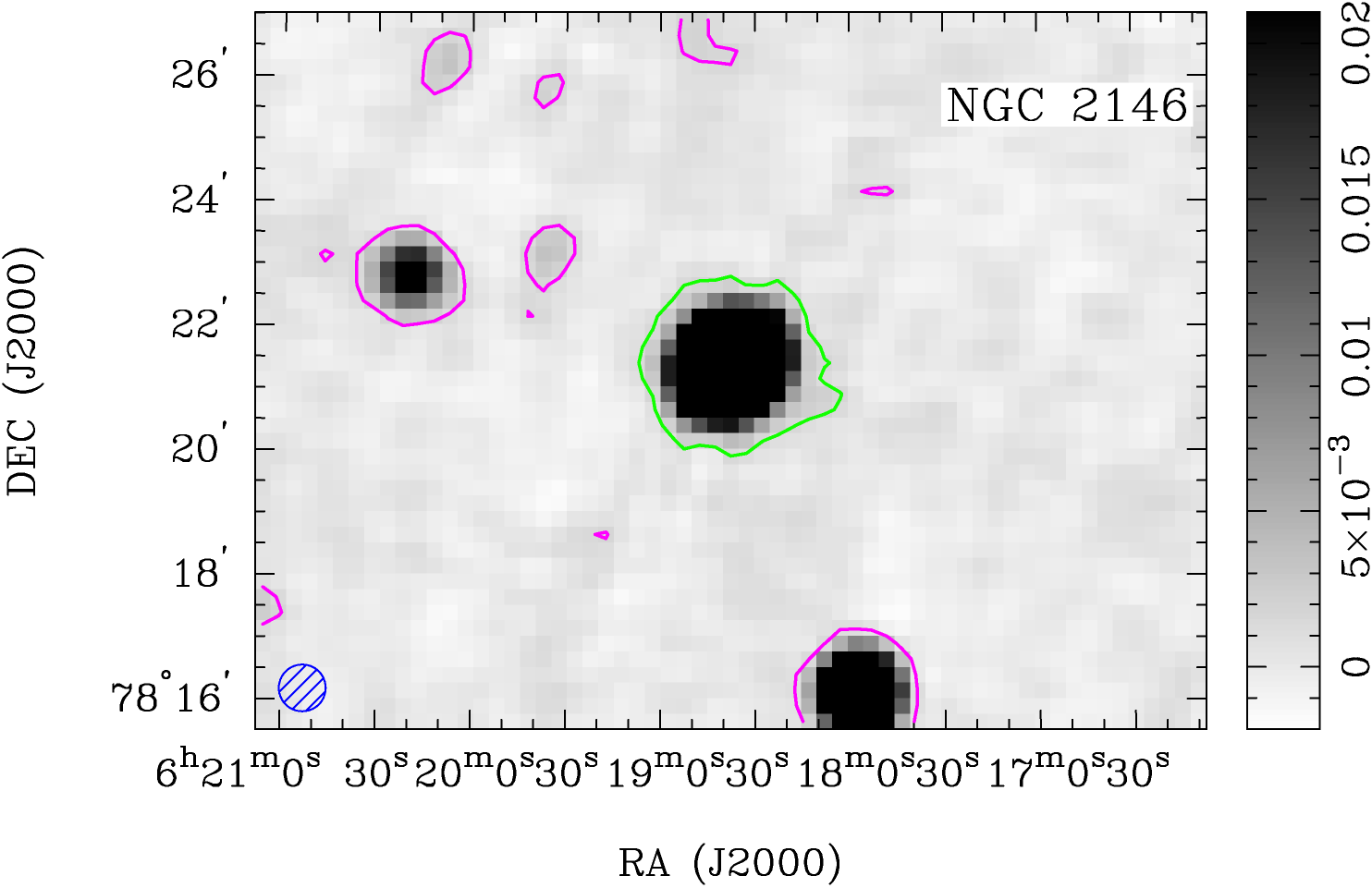}
      \includegraphics[angle=-90, width=0.38
  \textwidth]{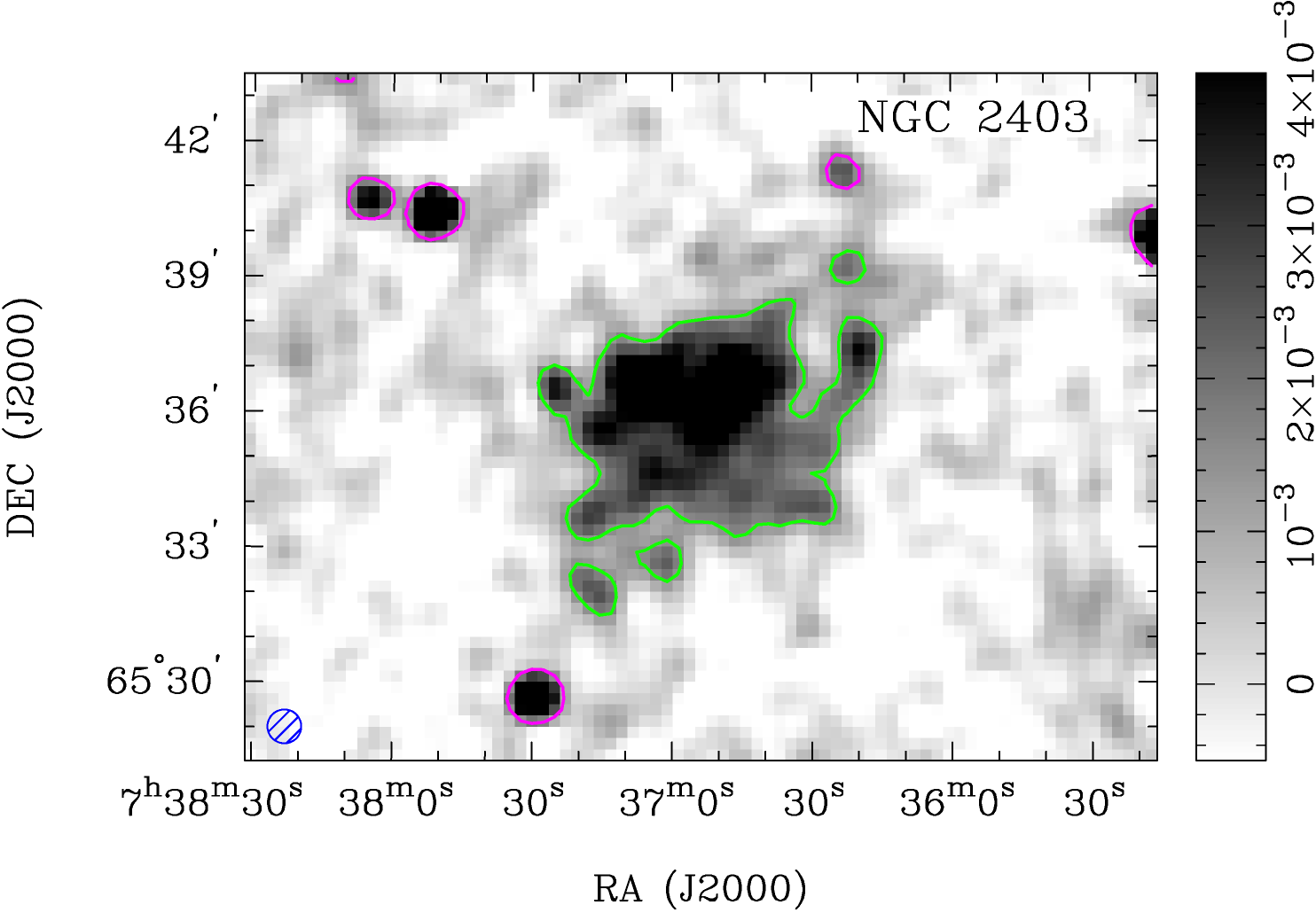} \\
  
    \includegraphics[angle=-90, width=0.38 \textwidth]{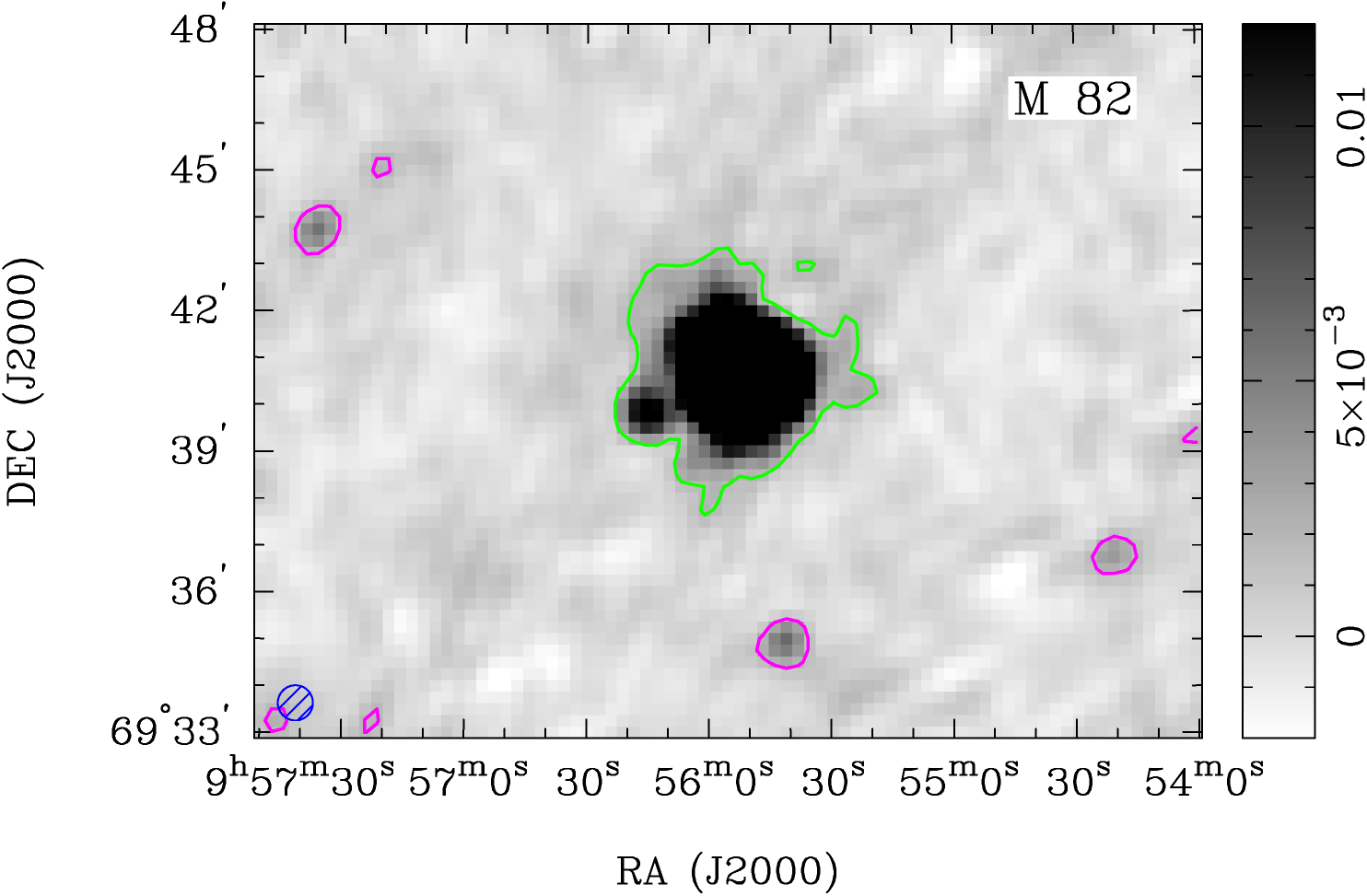}
      \includegraphics[angle=-90, width=0.38
      \textwidth]{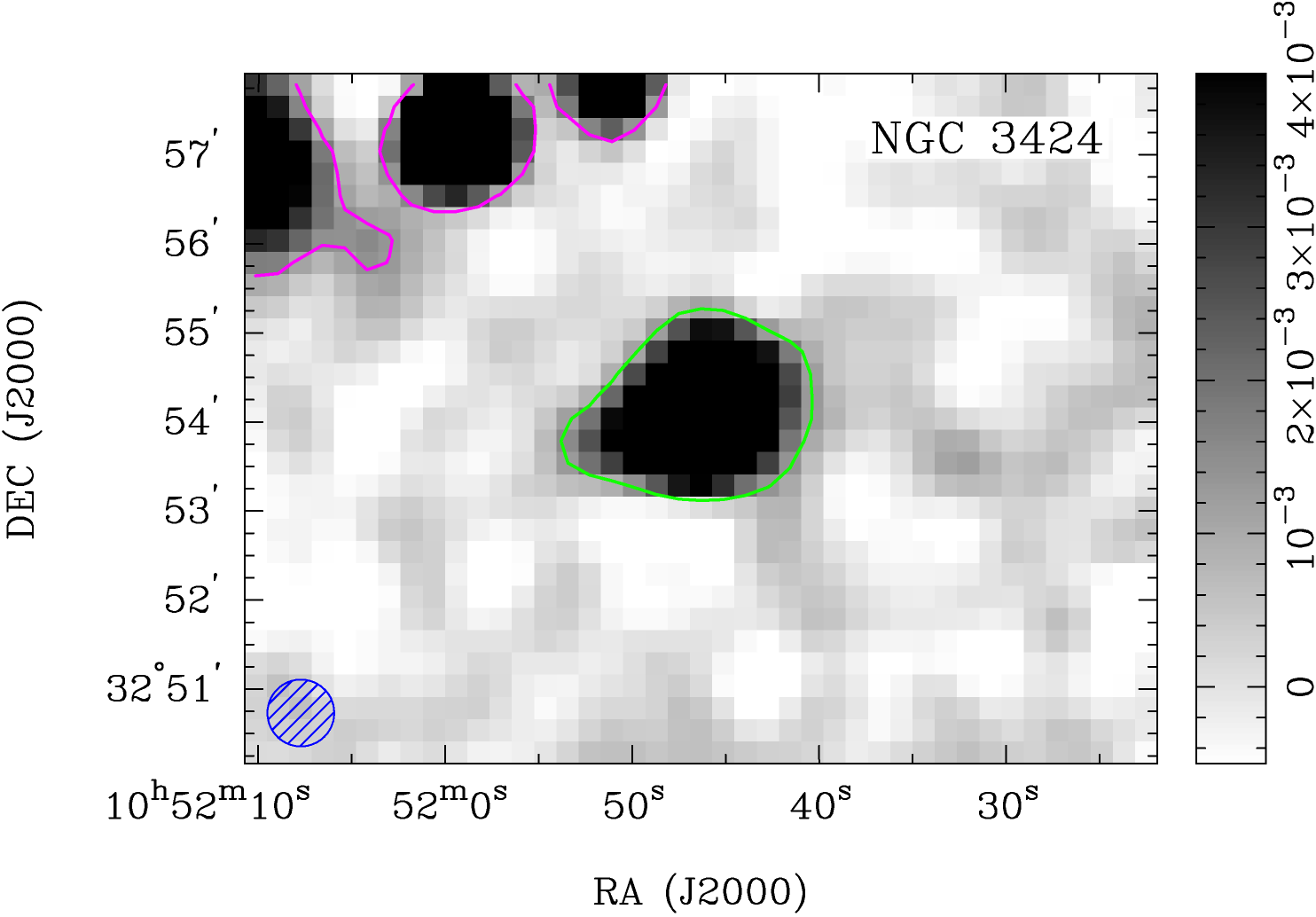}\\
  
  \includegraphics[angle=-90, width=0.38 \textwidth]{ 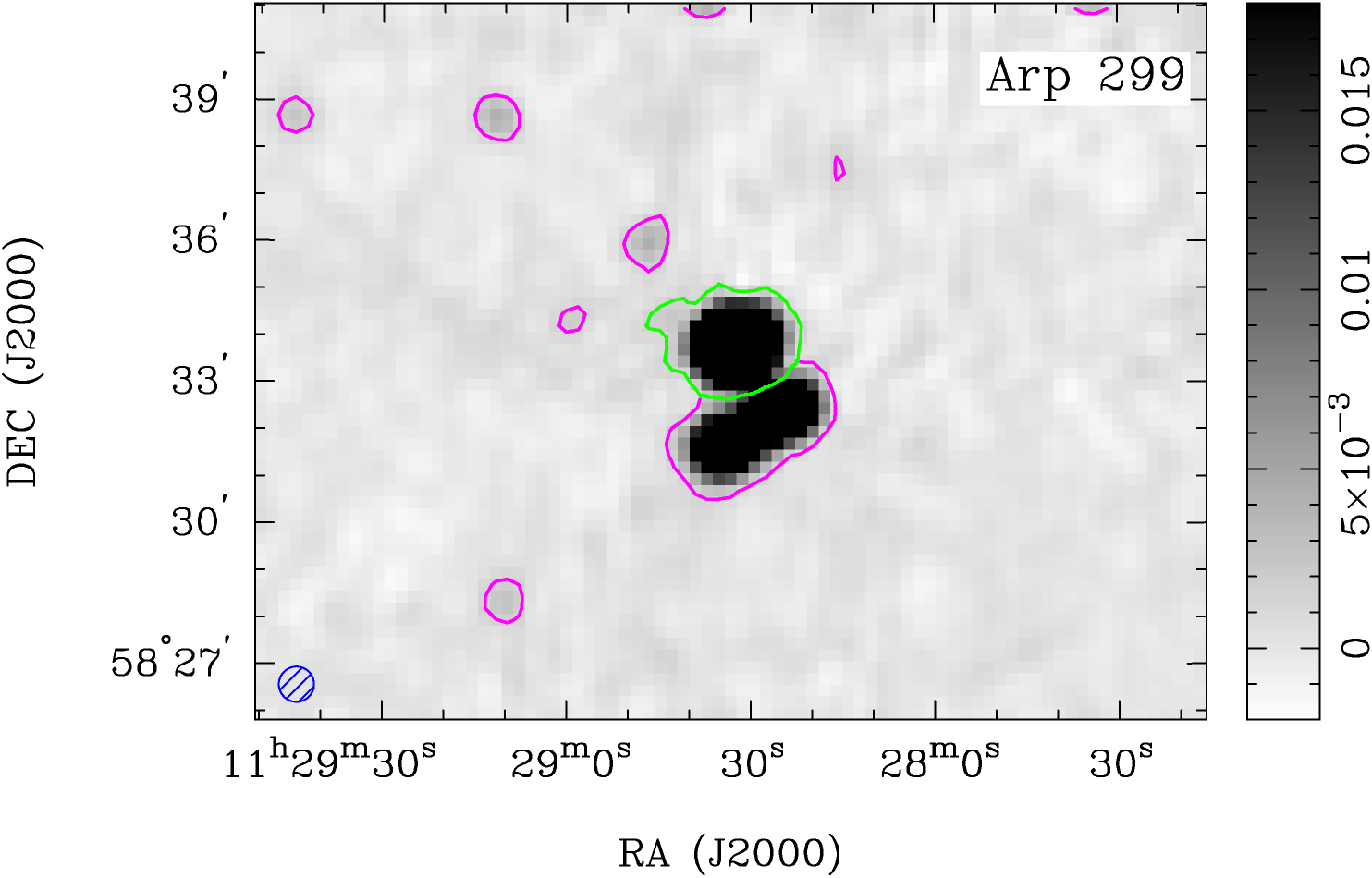}
      \includegraphics[angle=-90, width=0.38
  \textwidth]{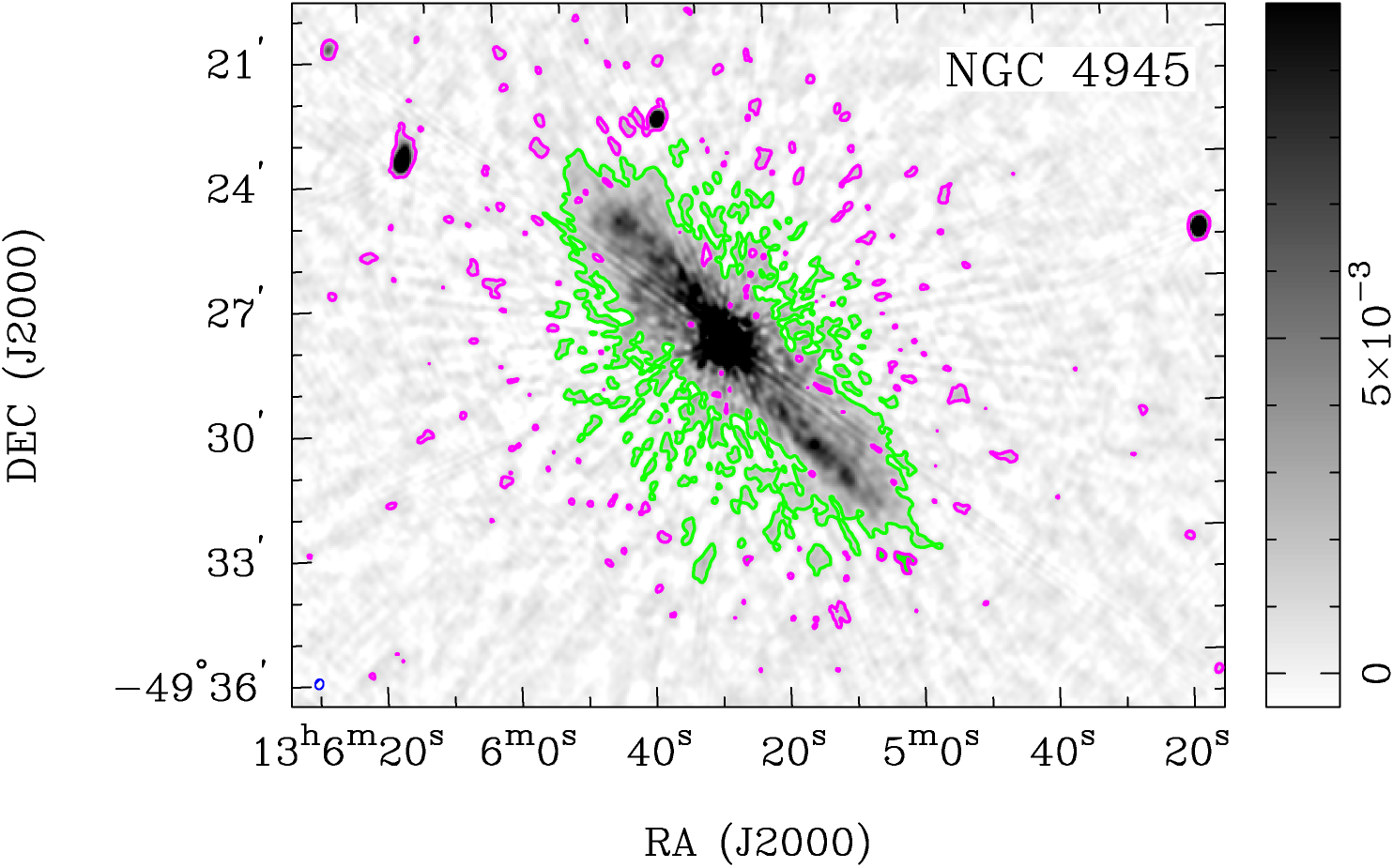}\\
  
      \includegraphics[angle=-90, width=0.39 \textwidth]{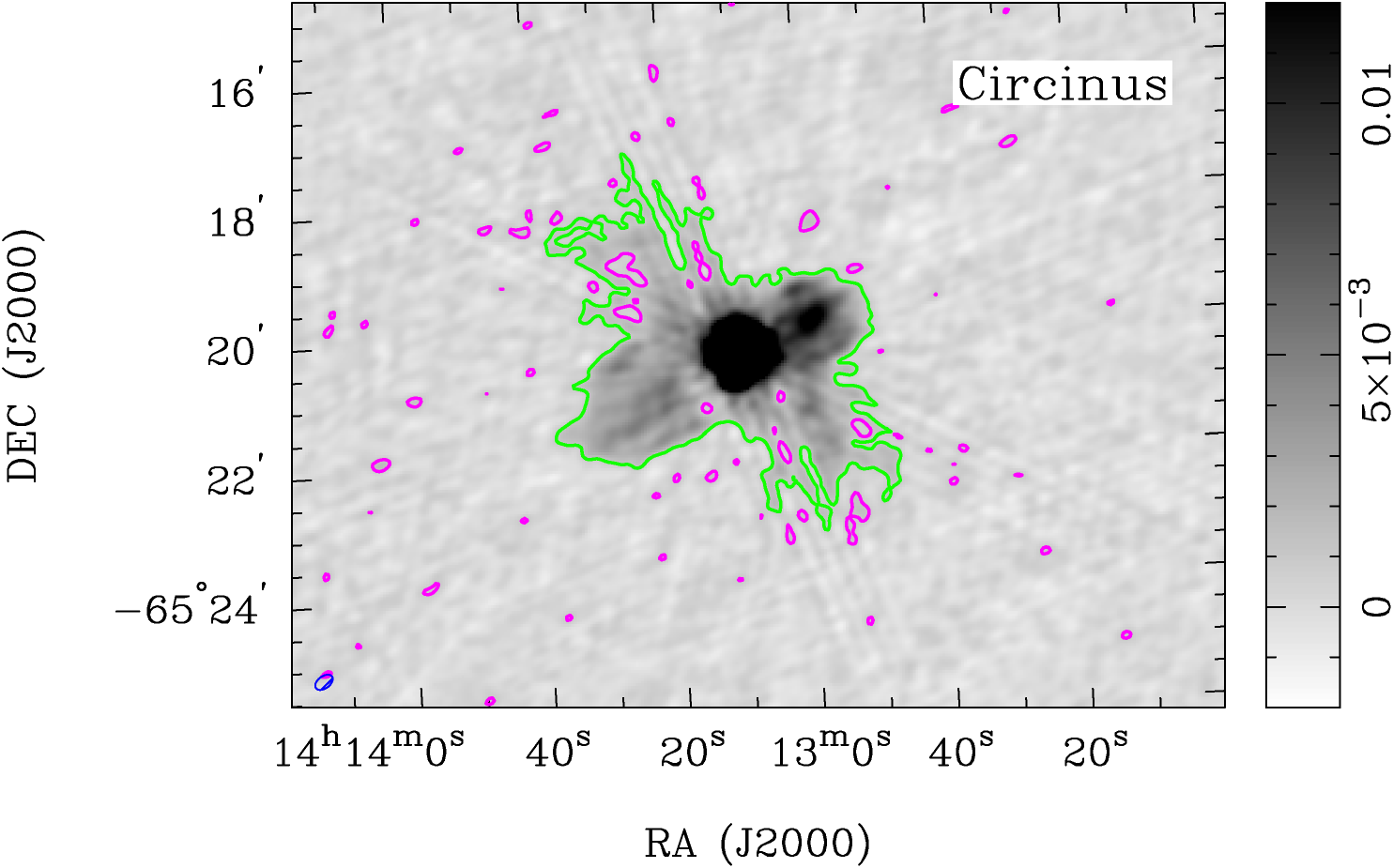}
    \includegraphics[angle=-90, width=0.38 \textwidth]{ 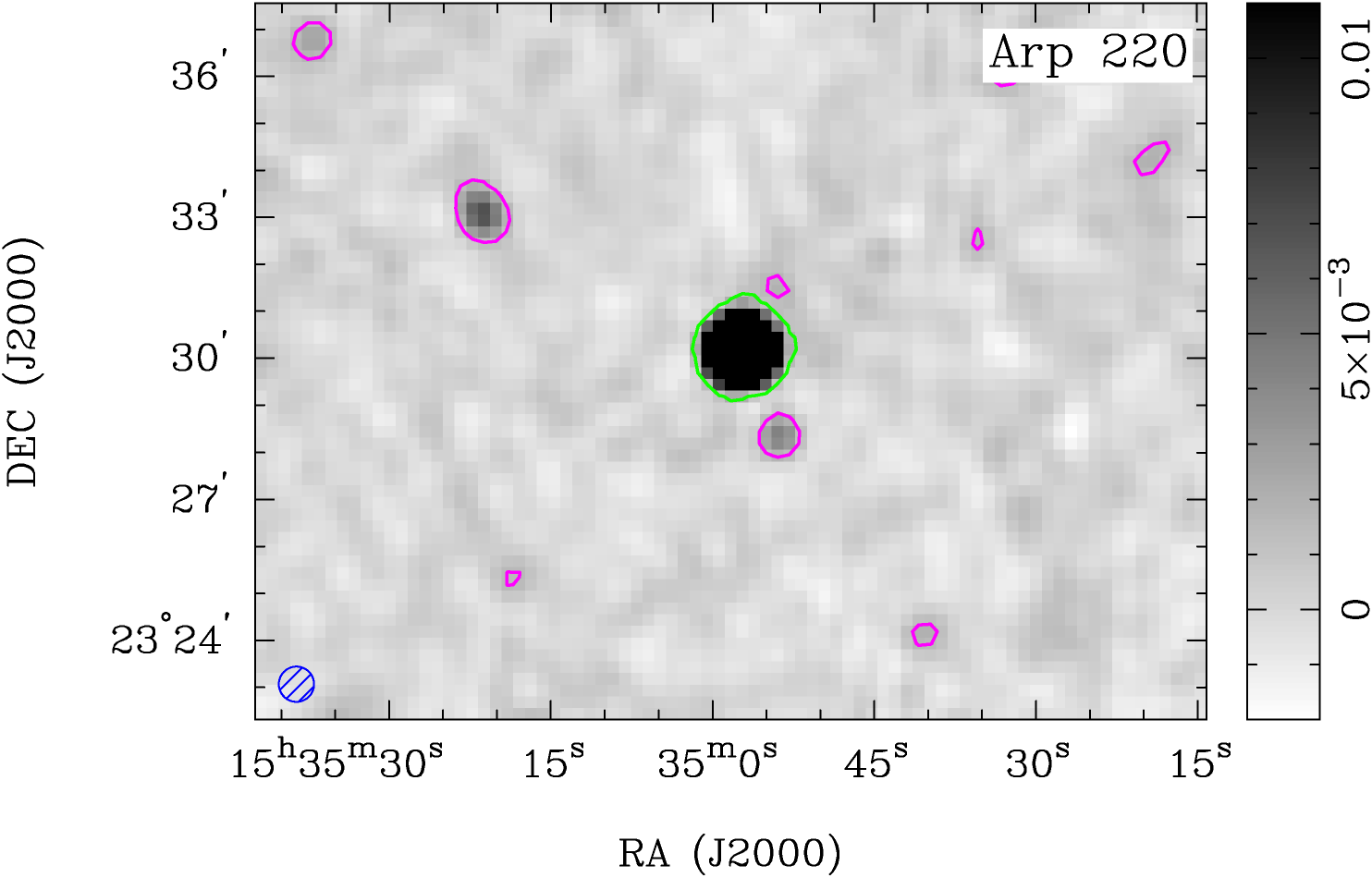}

\caption{Stokes-I continuum image at 1.4 GHz for each galaxy of our sample. The flux scale is in Jy~beam$^{-1}$. The magenta contours mark the 3-$\sigma$ level and the green solid contour the region selected for the flux integration.  The synthesised beam is shown with a blue circle at the bottom left corner of each image.}

  \label{imag_fluxes}
\end{figure*}

Here we present the Stokes-I continuum images at 1.4 GHz for the ten galaxies from which we extract the flux. In Fig.~\ref{Fig: images} we show the emission above 3 sigmas and the selected contours for the flux integration of each galaxy as explained in Sec.~\ref{sec:L1.4_vs_SFR}.

\section{Model details}
\label{ap:model}
In this appendix we specify the dependencies of the escape and cooling times with the free parameters of the model ($\beta, B_0, \rho$) and the SFR. 
We also show in Fig.~\ref{fig: SED} the modelled SEDs for the benchmark scenario at the two fixed SFRs presented in Sec.~\ref{subsec: General transport properties} ($\dot{M}_* = 0.1$ and $10\,\mathrm{M_\odot\,yr}^{-1}$).
Finally, we make a detailed derivation of the slope $\eta$ presented in Sec.~\ref{subsec: General transport properties}, using approximations in different ranges of SFRs.

\subsection{Cooling times}

Below we show the dependencies of the cooling times on the SFR and the free parameters of the model:

\begin{equation}
\label{Eq: Adv_dep}
\tau_{\rm adv}(\dot{M}) \propto \frac{\sqrt{\rho \, n}}{B} \propto \frac{\sqrt{\rho}}{ B_0 \, \dot{M}_*^{\beta-0.35}},  \, \, \, \dot{M}_* < 2 \, M_\odot \mathrm{yr}^{-1},
\end{equation}

\noindent
and it is constant for $ \dot{M}_* \geq 2 \, M_\odot\,\mathrm{yr}^{-1}$.

\begin{equation}
\tau_{D}(\dot{M}_*) \propto B^{\delta}\dot{M}_*^{\frac{1-\delta}{3}} \propto B_0^{1/3}\dot{M}_*^{\,\beta /3 + 2/9},
\end{equation}

\begin{equation}
\tau_{\mathrm{sync}}(\dot{M}_*) \propto \frac{1}{B^2} \propto \frac{1}{B_0^2 \, \dot{M}_*^{2 \beta}},
\end{equation}

\begin{equation}
\tau_{\mathrm{IC}}(\dot{M}_*) \propto \frac{1}{L_{\mathrm{IR}}} \propto \frac{1}{\dot{M}_*},
\end{equation}

\noindent  while $\tau_{\mathrm{brem}}$,  $\tau_{\mathrm{ion}}$,  $\tau_{\mathrm{pp}}$ are all $\propto n^{-1} \propto \dot{M}_*^{-0.71}$.

\subsection{Model SEDs}

\begin{figure*}
    \centering
    \includegraphics[angle=-90, width=0.494\textwidth,]{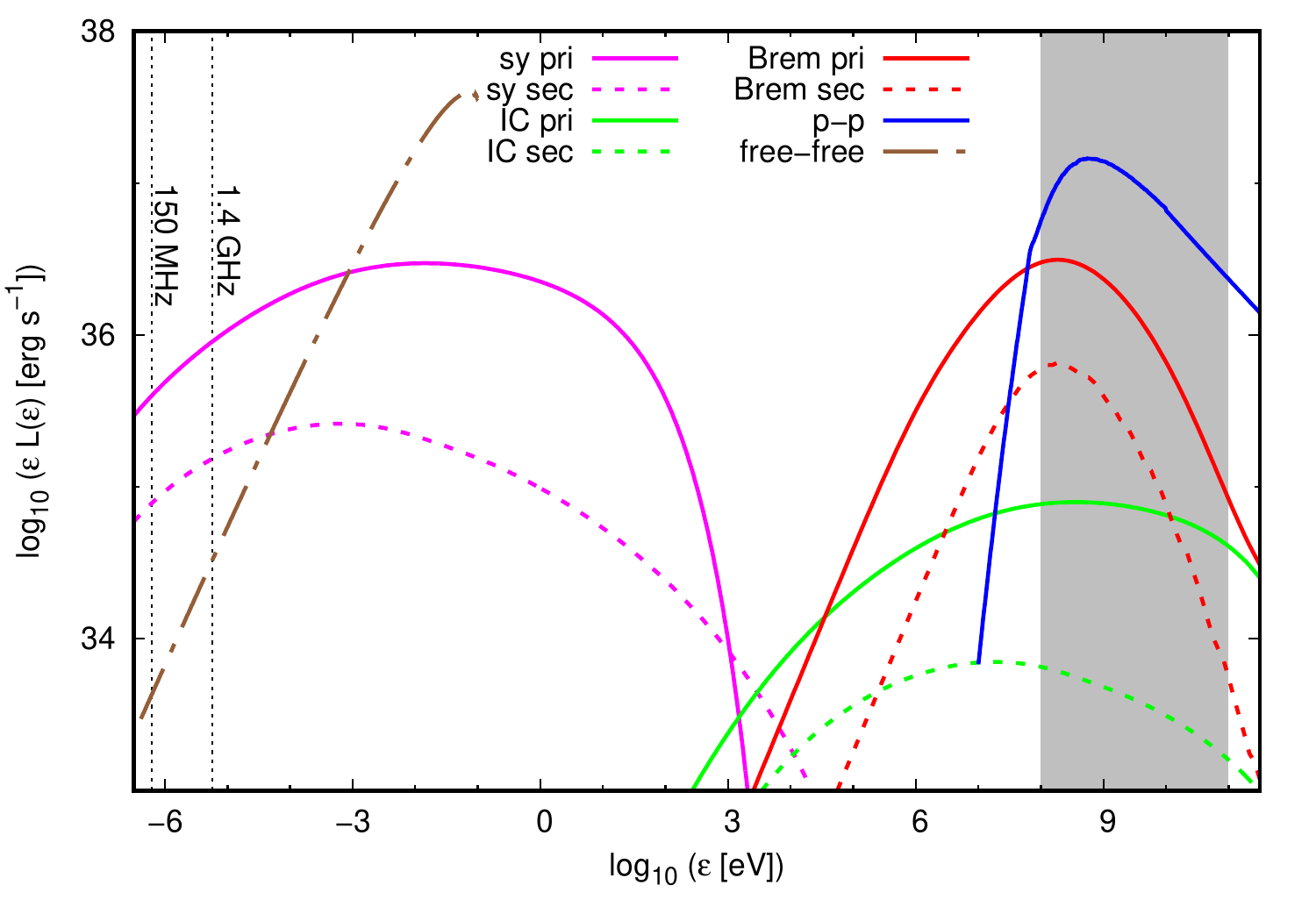}
    \includegraphics[angle=-90, width=0.494\textwidth]{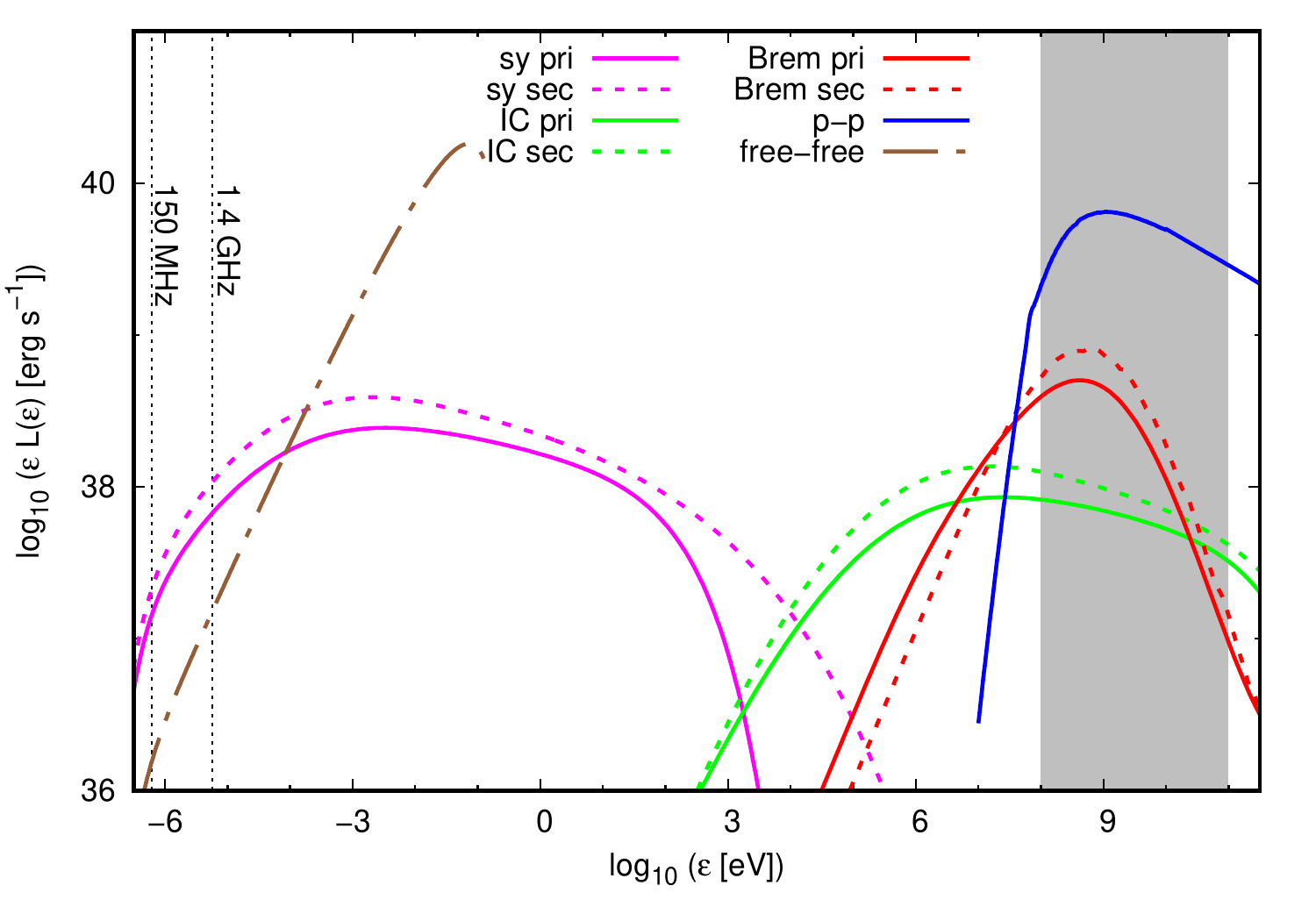}
    \caption{SEDs computed from scenario 0, for typical cases of low ($\dot{M}_*$ = 0.1 $M_{\odot}/\mathrm{yr}$, left) and high ($\dot{M}_*$ = 10 $M_{\odot}/\mathrm{yr}$, right) SFRs. The coloured lines (solid for the primary CRs and dashed for the secondary) are the individual contributions of different non-thermal radiative mechanisms (the colour code is the same as in Fig~\ref{fig: cooling}). The dashed brown line is the thermal free-free emission. The grey shaded region represents the \textit{Fermi-LAT} energy range and the vertical dotted black lines show the studied radio frequencies (150 MHz and 1.4 GHz).}
    \label{fig: SED}
\end{figure*}

We show the modelled SEDs for $\dot{M}_*$ = 0.1 $\rm M_{\odot} yr^{-1}$ on the left panel of Fig.~\ref{fig: SED}. The dominant contribution at 150 MHz and 1.4 GHz comes from the synchrotron emission of primary electrons. In the case of $\dot{M}_*$ = 10 $\rm M_{\odot} yr^{-1}$, that we show on right panel of Fig.~\ref{fig: SED}, the dominant contribution comes instead from synchrotron emission of secondary electrons. In both cases, the free-free thermal component becomes important at higher frequencies.
In the \textit{Fermi} energy range the emission is dominated by hadronic emission for both SFRs.

\subsection{The slope of the $L_{1.4\mathrm{GHz}}$--SFR relation}

In order to understand the behaviour of the slope $\eta$ of Eq.\ref{eq: Lsyn_dep} in the benchmark scenario, we approximate the electron distribution as a power-law ($N(E) = N_0~E^{-p}$) in the electron-energy range producing the dominant synchrotron contribution to radiation at 1.4~GHz ($\mathcal{E}_{1.4}$). This range comprises between $\sim 1$~GeV and $\sim 5$~GeV for the benchmark scenario. Under this assumption, the synchrotron luminosity at a given frequency can be estimated roughly as $L_{\rm syn}^{1.4} \propto N_0~B^{(p+1)/2} $ \citep{Ghisellini2013}. For simplicity  we focus on two typical values at low and high SFRs, where we can assume that electrons lose their energy through a single dominant process near $\mathcal{E}_{1.4}$.

At low SFR $L_{\rm syn}^{1.4}$ is shaped by primary synchrotron emission, and near $\dot{M}_* = 0.1\rm M_{\odot} \rm yr^{-1}$ the electrons distribution is manly dominated by diffusion at $\mathcal{E}_{1.4} \sim 4\,\mathrm{GeV}$ (see upper left panel on Fig~\ref{fig: cooling}). Then $p=\alpha+\delta$ and $N_{0, \rm pri} \propto \dot{M_*} \tau_{D}$, which results in $L_{\rm syn}^{1.4} \propto  \dot{M_*}^{1.8}.$
At high SFR the slope of the primary electrons subdominant contribution can be estimated in a similar way taking into account that Bremsstrahlung dominates cooling. Then $p=\alpha$, $N_{0, \rm pri}\propto \dot{M_*} \tau_{\rm brem}$ and $L_{\rm syn}^{1.4} \propto  \dot{M_*}^{0.8}$. 

The emission of secondary electrons is dominant in $L_{\rm syn}^{1.4}$ at high SFR. The normalisation of their particle distribution depends on the proton distribution roughly as $N_{0,\rm sec}~\propto~n \, Q_{0,p} \tau_{p} \tau_{e} \propto \dot{M_*} \dot{M_*}^{0.7} \tau_{p} \tau_{e}$ (see section~\ref{SubsecSec: Seconday}), where $\tau_{p}$ and $\tau_{e}$ are the dominant loss times for protons and electrons, respectively.

Near $\dot{M}_* \sim 10 \, \rm M_{\odot} \rm yr^{-1}$ proton-proton interactions and advection dominate the proton losses, and Bremsstrahlung controls the electron cooling near $\mathcal{E}_{1.4}$ (see lower right panel on Fig~\ref{fig: cooling}). On the one hand if we assume that proton-proton interactions dominate alone, then $p=\alpha$, $N_{0, \rm sec} \propto \dot{M_*}^{1.7} \tau_{\rm pp}\tau_{\rm brem}$ and $L_{\rm syn}^{1.4} \propto \dot{M_*}^{0.8}$. On the other hand if we assume that advection domains alone, then $N_{0, \rm sec} \propto \dot{M_*}^{1.7} \tau_{\rm adv}\tau_{\rm brem}$ where $L_{\mathrm{syn}} \propto  \dot{M_*}^{1.5}$. 
The combined effect of these two losses returns a dependence in SFR between these last two values.

\section{Hadronic processes}
\label{Appendix-Kelner}

For the sake of completeness we report here the functions we adopted in the computation of all hadronic byproducts.
\begin{eqnarray}
F_{\gamma}(x,E_p)= B_{\gamma} \frac{d}{dx} \Big[ {\rm ln}(x) \Big( \frac{1-x^{\beta_{\gamma}}}{1+k_{\gamma}x^{\beta_{\gamma}}(1-x^{\beta_{\gamma}})} \Big)^4 \Big] = \\
B_{\gamma} \frac{{\rm ln}(x)}{x} \Big( \frac{1-x^{\beta_{\gamma}}}{1+k_{\gamma}x^{\beta_{\gamma}}(1-x^{\beta_{\gamma}})} \Big)^4 \times \nonumber \\
\Big[ \frac{1}{{\rm ln}(x)} - \frac{4 \beta_{\gamma} x^{\beta_{\gamma}}}{1-x^{\beta_{\gamma}}} - \frac{4 k_{\gamma} \beta_{\gamma} x^{\beta_{\gamma}} (1-2x^{\beta_{\gamma}})}{1+k_{\gamma}x^{\beta_{\gamma}}(1-x^{\beta_{\gamma}})} \Big] \nonumber
\end{eqnarray}
where $x=E_{\gamma}/E_p$ and 
\begin{eqnarray}
B_{\gamma}= 1.30 + 0.14 \, L + 0.011 \, L^2 \\
\beta_{\gamma} = \frac{1}{1.79 + 0.11 \, L + 0.008 \, L^2}  \\
k_{\gamma} = \frac{1}{0.801 + 0.049 \, L + 0.014 \, L^2}
\end{eqnarray}
and where $L= {\rm ln}(E_p/1 \, \rm TeV) $.
\begin{equation}
    F_e(x,E_p) = B_e \, \frac{[1+k_e({\rm ln} x)^2]^3}{x(1+0.3/x^{\beta_e})} [- {\rm ln}(x)]^5,
\end{equation}
where 
\begin{eqnarray}
B_{e}= \frac{1}{69.5 + 2.65 \, L + 0.3 \, L^2} \\
\beta_{e} = \frac{1}{(0.201 + 0.062 \, L + 0.00042 \, L^2)^{1/4}}  \\
k_{e} = \frac{0.279 + 0.141 \, L + 0.0172 \, L^2}{0.3 + ( 2.3 + L^2)^2} 
\end{eqnarray}

We also report the function $\tilde{f}_e$: 
\begin{equation}
    \tilde{f}_e(x)= f_{\nu_{\mu}^{(2)}}(x) = g_{\nu_{\mu}}(x) \, \Theta(x-r) + [h_{\nu_{\mu}}^{(1)}(x)+h_{\nu_{\mu}}^{(2)}(x)] \Theta(r-x) 
\end{equation}
where $r=1-\lambda=(m_{\mu}/m_{\pi})^2=0.573$,
\begin{eqnarray}
g_{\nu_{\mu}}(x)= \frac{3-2r}{9(1-r)^2}(9x^2-6 \, {\rm ln} \, x - 4x^3 -5) \\
h^{(1)}_{\nu_{\mu}}(x)= \frac{3-2r}{9(1-r)^2}(9r^2-6 \, {\rm ln} \, r - 4r^3 -5) \\
h^{(2)}_{\nu_{\mu}}(x)= \frac{(1+2r)(r-x)}{9 \,r^2} \times [9\,(r+x)- 4 \,(r^2+rx+x^2)],
\end{eqnarray}
$\Theta$ is the Heaviside step function.

\end{appendix}

\end{document}